\documentclass[twocolumn,twoside,slac_two]{revtex4}
\usepackage{bm}
\usepackage{graphicx}
\usepackage{fancyhdr}
\usepackage[dvips]{hyperref}

\newcommand{\rpv}{\mbox{$\not \hspace{-0.10cm} R$}}
\newcommand{\MET}{\mbox{$\not \hspace{-0.11cm} E_T$}}
\newcommand{\lle}{\mbox{$LL\bar{E}$}}
\newcommand{\lqd}{\mbox{$LQ\bar{D}$}}

\newcommand{\neutralino}{\mbox{$\tilde{\chi}_1^0$}}
\newcommand{\chargino}{\mbox{$\tilde{\chi}^\pm_1$}}

\newcommand{\lesim}{\mbox{$\;\raisebox{-1mm}{$\stackrel{\textstyle<}{\sim}$}\;$}}

\begin{document}

\title{Searches for New Phenomena at the Tevatron and at HERA}

\author{Arnd Meyer}
\affiliation{\mbox{III.\,Phys.\,Inst.\,A}, RWTH Aachen, 52074 Aachen, Germany \\
E-mail: meyera@physik.rwth-aachen.de} 
%

\begin{abstract}

  Recent results on searches for new physics at Run II of the Tevatron and highlights
  from HERA are reported. The searches cover many different final states and a wide range
  of models. All analyses have at this point led to negative results, but some interesting
  anomalies have been found.

\end{abstract}

\maketitle

\thispagestyle{fancy}

\section{Introduction}

  For all of the deficits of the standard model (SM) that we know about since many years --
  be it the non-unification of couplings at a high scale, the quadratic divergences in the
  loop corrections to the Higgs boson mass, or the lack of a decent dark matter candidate --
  a large number of solutions has been proposed. We know that within the standard model, the
  $W_L W_L$ scattering amplitude violates the unitarity bound at a center of mass energy
  $\simeq 1.7$~TeV \cite{unit}, and one solution to this problem is offered by the Higgs
  mechanism \cite{higgs}, through the introduction of a massive scalar particle. To
  successfully address the $W_L W_L$ scattering amplitude problem, the Higgs boson mass is
  constrained to $m_H\lesim 1$~TeV, and if fermions acquire their masses through coupling to
  the Higgs boson, then  $m_H \lesim 200$~GeV is required \cite{lepew}. If the Higgs boson
  doesn't exist, some other form of new physics must be present at the TeV scale to prevent
  the $W_L W_L$ scattering amplitude from violating the unitarity bound.

  The most popular models of new physics involve without doubt supersymmetry. However,
  supersymmetry doesn't explain the number of fermion generations, or their mass spectrum and
  charges. In this talk, recent results from the Tevatron for searches for manifestations of new
  physics are reported, in the areas of supersymmetry, extra gauge bosons, leptoquarks, large
  extra dimensions, quark and lepton compositeness, the Higgs sector, and a few signature based
  searches. In addition, selected highlights from signature based searches at HERA are presented.

\begin{figure}[bth]
    \includegraphics[height=.27\textheight]{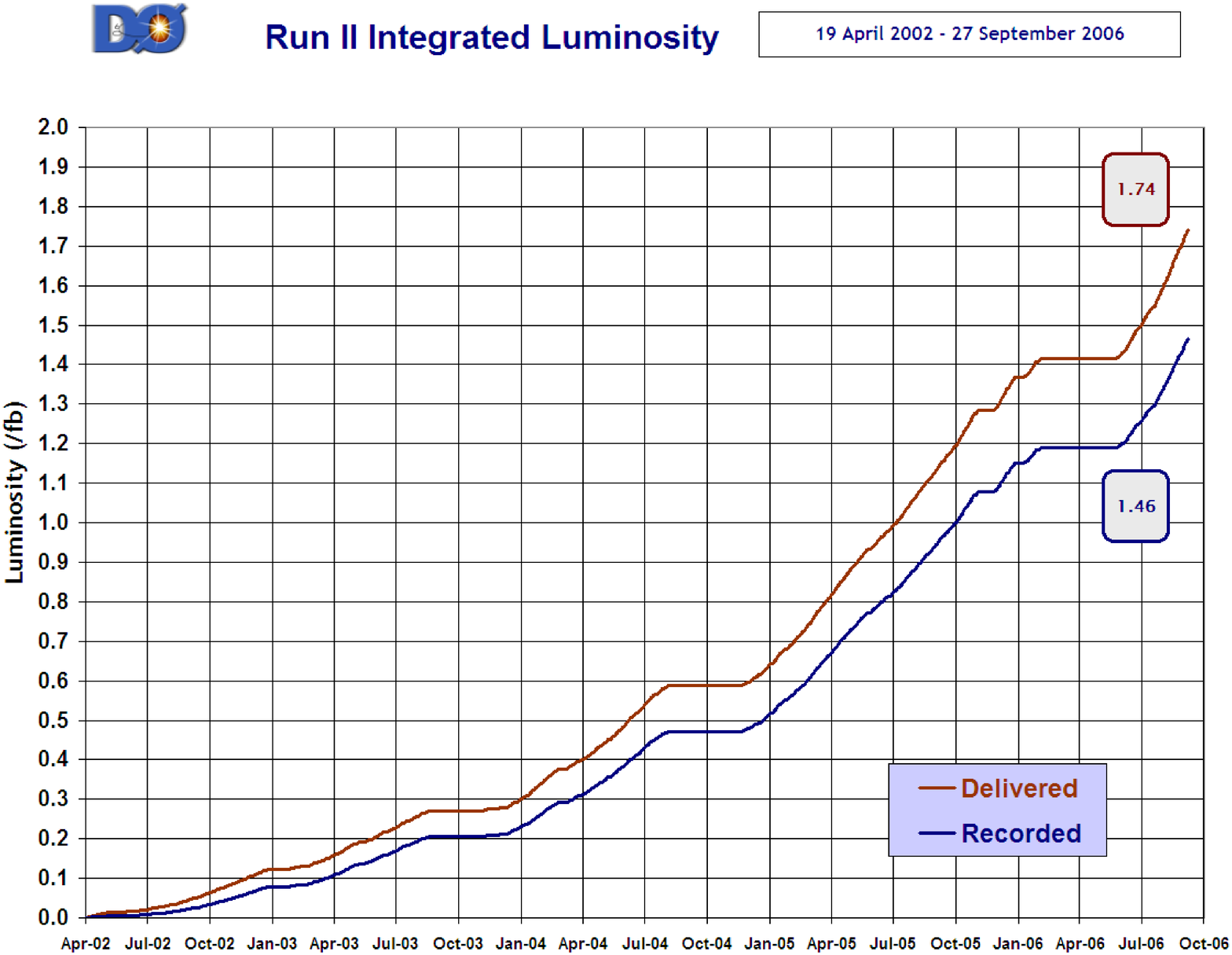}
    \caption{Integrated luminosity delivered by the Tevatron and recorded by
             D\O\ during Run II.}
    \label{d0lumi}
  \end{figure}

  The Tevatron is a proton-antiproton collider with a collision energy of 1.96~TeV in the
  center of mass system. It is situated at Fermilab near Chicago. Run IIa has ended in February of
  2006 with a dataset corresponding to an integrated luminosity of $\simeq 1.3$~fb$^{-1}$
  per experiment. This represents about 10 times the statistics collected in Run I. Run IIb
  has started in June of 2006 with the goal to reach at least 4, but possibly 8~fb$^{-1}$ by
  the year 2009. The two experiments D\O\ and CDF are by now well understood in their
  capabilities to detect and identify electrons, photons, muons, taus, jets of light and heavy
  flavours, and missing transverse energy \MET. The current account of delivered and recorded
  luminosity by D\O\ is shown in Fig.~\ref{d0lumi}. Only recent results based on an integrated
  luminosity of at least 0.3~fb$^{-1}$ are reported here. Details can be found on the
  corresponding experiment web sites \cite{prelim}.

\begin{figure}[bth]
    \includegraphics[height=.26\textheight]{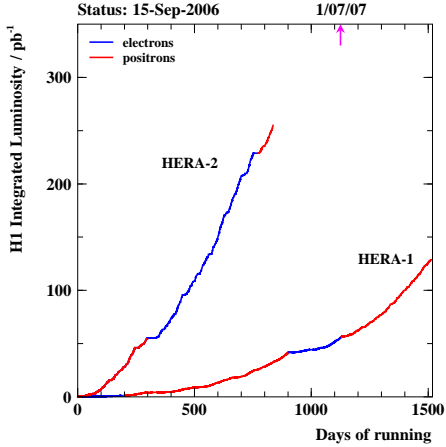}
    \caption{Integrated luminosity accumulated by the H1 detector at HERA \cite{h1lumiref}.}
    \label{h1lumi}
  \end{figure}

  The HERA electron(positron)-proton collider at DESY in Hamburg has been delivering
  luminosity since 1992, at a center of mass energy of up to 319~GeV. HERA is currently in
  its ``Run II'' (HERA-2) as well, which started in 2002 and is expected to be completed in 2007.
  The total integrated luminosity accumulated by the two colliding beam experiments H1 and
  ZEUS for the analyses shown here is up to $\simeq 300$~pb$^{-1}$, with another
  substantial increase of the data set expected for the entire HERA data set, as can be
  seen for H1 in Fig.~\ref{h1lumi}.

\section{Isolated Leptons at HERA}

\begin{figure}[bth]\setlength{\unitlength}{1cm}
    \begin{picture}(8.6,3.2)(0.0,0.0)
      \put(-5.3,8.9) {\includegraphics[height=.85\textheight,angle=270]{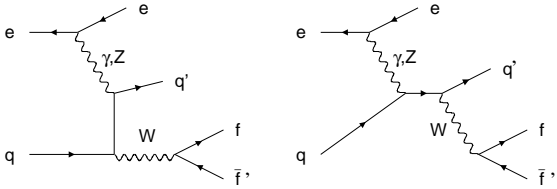}}
    \end{picture}
    \caption{Main leading order Feynman diagrams for the process $ep \rightarrow e W X$ \cite{zeusiso2006}.}
    \label{wprod}
  \end{figure}

  Soon after turning on HERA, the H1 experiment reported \cite{h1_1994} in 1994, after analyzing
  the first 4~pb$^{-1}$ of data, the observation of an event with an isolated muon recoiling
  against a hadronic system, both of high transverse momentum $p_T$. In addition, substantial missing
  transverse energy was reconstructed. The dominant SM process leading to such a final state is
  photoproduction of $W$ bosons with a subsequent leptonic decay of the $W$: $e^+ p \rightarrow
  e^+ W^+ X \rightarrow e^+ \mu^+ \nu X$ (Fig.~\ref{wprod}), where the positron escapes detection
  and the neutrino leads to reconstructed \MET. However, the total cross section for this
  reaction is only $40$~fb. In the following years, H1 has reported on the observation of more
  such events, in excess of the SM expectation, but not statistically significant to claim new
  physics.

  These analyses have inspired a large number of possible interpretations in terms
  of physics beyond the standard model -- to name a few, leptoquarks, excited fermions,
  supersymmetry with $R$-parity violation, or single top production via flavour changing
  neutral currents.

  \subsection{High $\mathbf p_T$ Leptons and Missing $\mathbf E_T$}

\begin{figure}[bth]
      \includegraphics[height=.25\textheight]{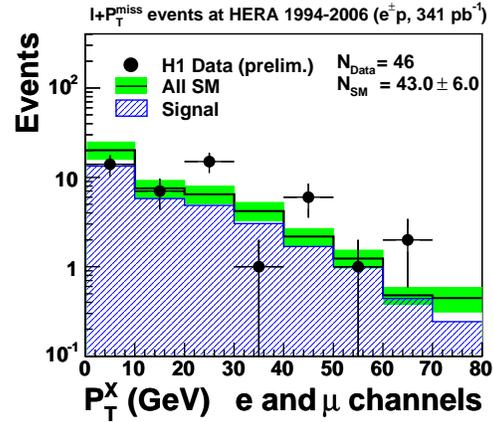}
      \caption{The hadronic transverse momentum distribution in the H1 search for isolated
      lepton events \cite{h1iso2006}. Electron and muon channels are combined. The SM expectation
      (histogram) is dominated by processes with genuine isolated leptons and missing $E_T$ (``Signal''),
      which in turn is dominated by real $W$ production.}
      \label{h1leptonscomb}
    \end{figure}

    The H1 collaboration has now updated the analysis with all data collected until the Summer of
    2006, corresponding to 341~pb$^{-1}$ of data \cite{h1iso2006}. The event selection requires as
    before a high $p_T$ isolated electron or muon, and substantial missing transverse momentum.
    The distribution of the hadronic transverse momentum $p_T^X$, determined from all
    reconstructed particles excluding identified isolated leptons, for all data combined is shown
    in Fig.~\ref{h1leptonscomb}, with a fair agreement of data and SM expectation.

\begin{figure}[bth]\setlength{\unitlength}{1cm}
      \begin{picture}(8.6,4.5)(0.0,0.0)
	\put(-3.1,7.7) {\includegraphics[height=.65\textheight,angle=270]{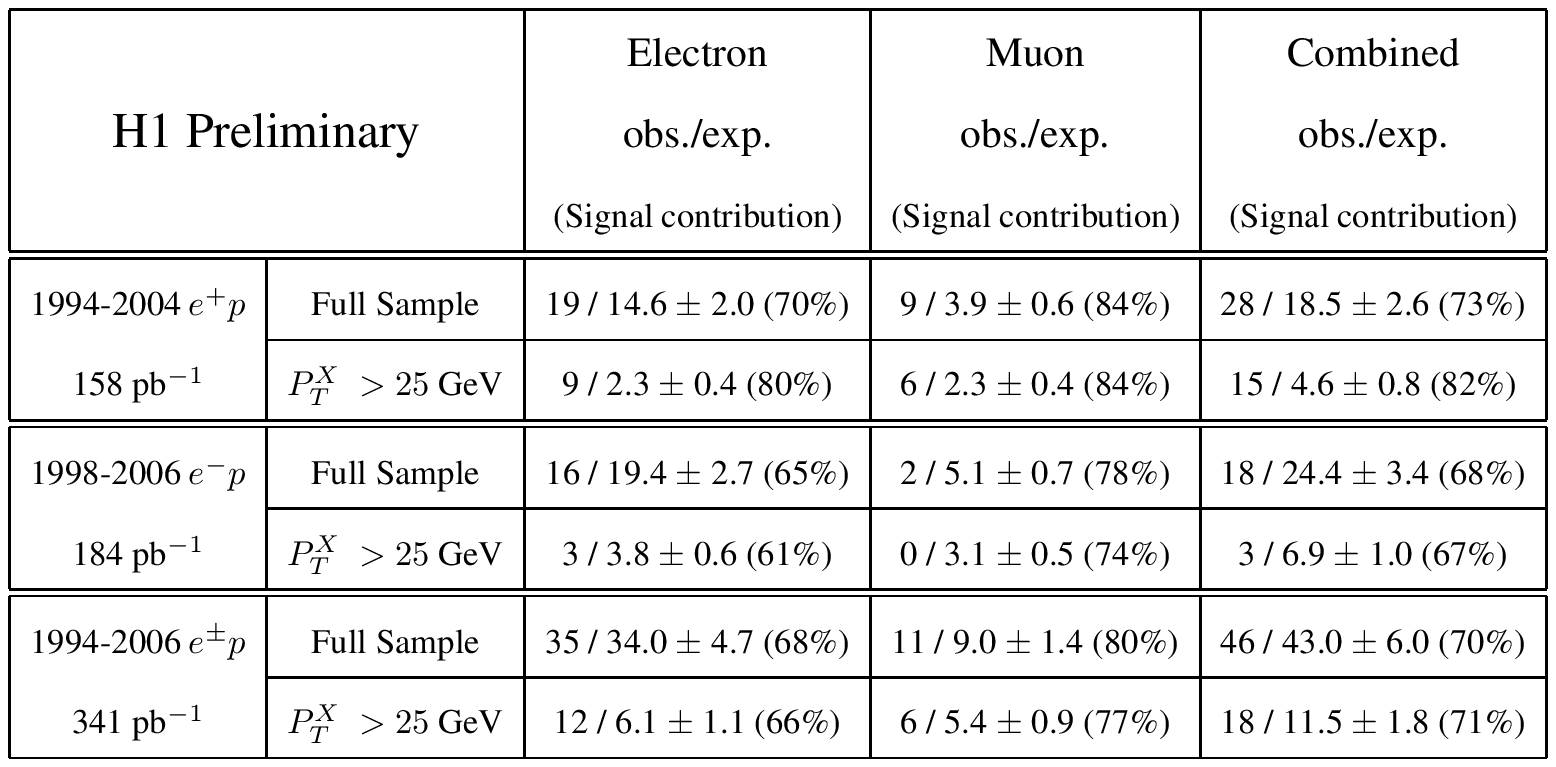}}
      \end{picture}
      \caption{Summary \cite{h1iso2006} of the H1 results of searches for events with isolated electrons or muons and
	 missing transverse momentum for different data sets: $e^+p$, $e^-p$, and all data; $e$ and $\mu$ channel as well
	 as combined; with and without the requirement of large hadronic transverse momentum $p_T^X$. The number of
	 observed events is compared to the SM prediction. The signal component of the SM expectation, dominated by
	 real $W$ production, is given as a percentage in parentheses.}
      \label{h1leptonstable}
    \end{figure}

    At large hadronic transverse momentum $P_T^X > 25$~GeV, a total of 18 events are observed
    compared to a SM prediction of $11.5\pm 1.8$. As can be seen from Table \ref{h1leptonstable},
    the excess is, within statistics, observed in the $e^+p$ data only. The probability for the
    SM expectation to fluctuate to the observed number of events or more in the high $P_T^X$ domain
    for all data is 6.7\%, compared to 0.15\%\ for the HERA I data (the majority of which is $e^+p$
    data). For the $e^+p$ data alone, this probability is 0.03\%.

\begin{figure}[bth]\setlength{\unitlength}{1cm}
      \begin{picture}(8.6,7.5)(0.0,0.0)
	\put(-0.3,0.3) {\includegraphics[height=.128\textheight]{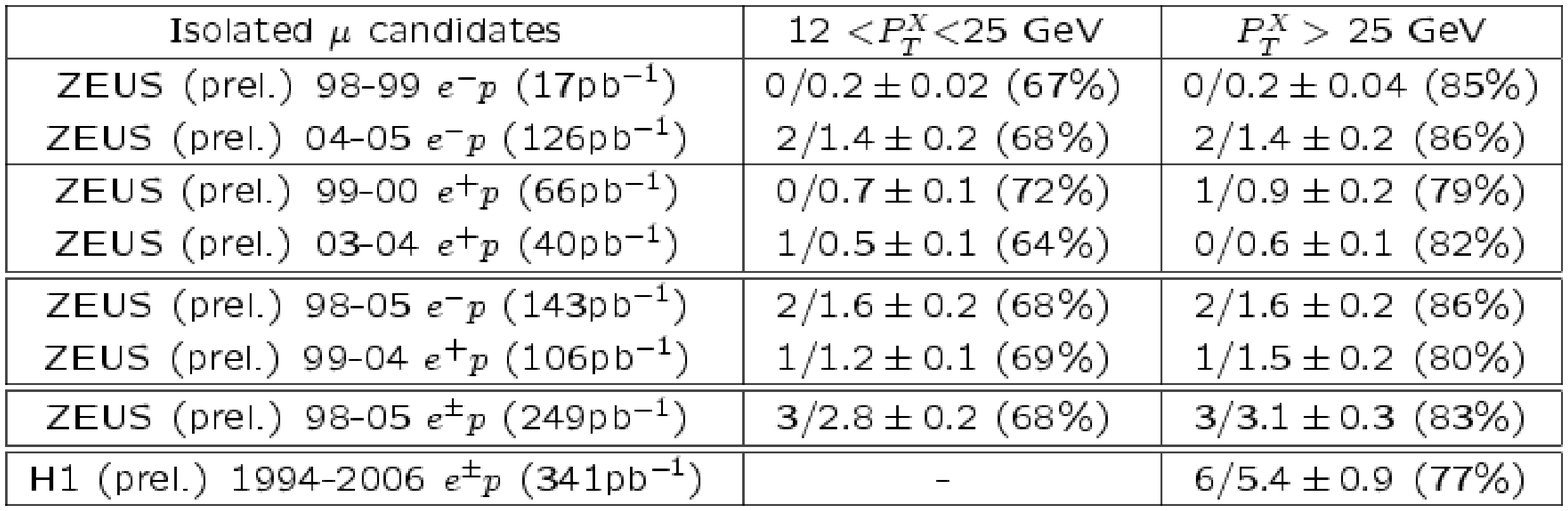}}
	\put(-0.2,4.0) {\includegraphics[height=.118\textheight]{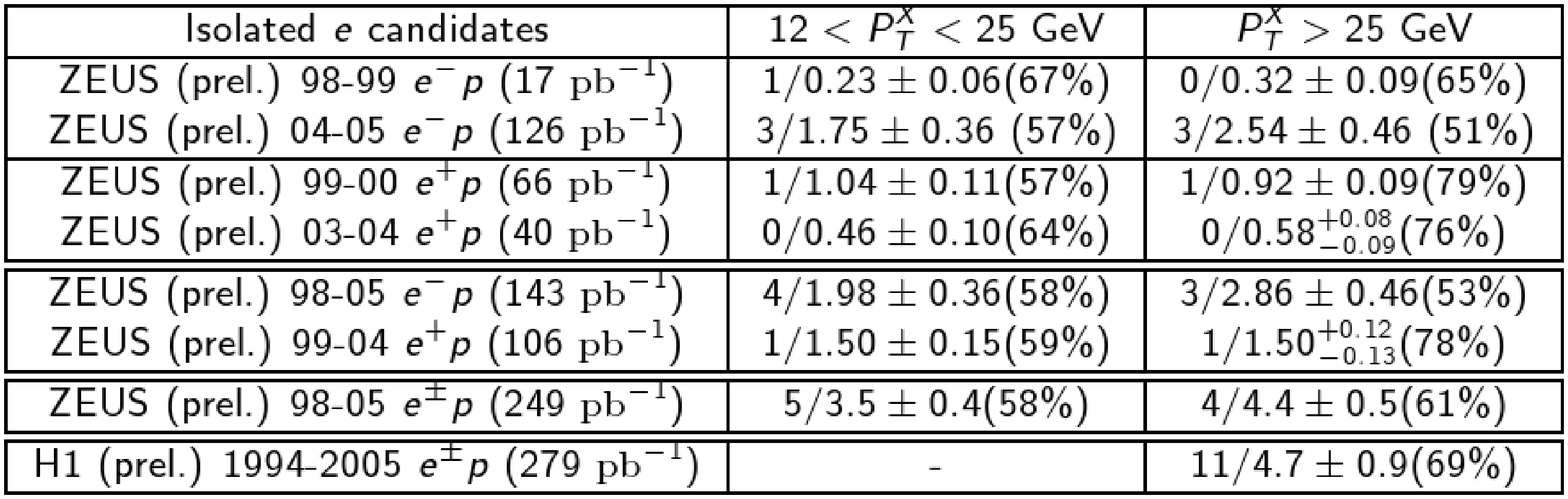}}
      \end{picture}
      \caption{Summary of the results of searches for events with isolated electrons (top)
       or muons (bottom) and missing transverse momentum at HERA, as shown in \cite{zeusiso2006}.
       The number of observed events is compared to the SM prediction (observed/expected). The signal
       component of the SM expectation ($W$ production) is given as a percentage in parentheses. Only
       the H1 results directly comparable to the ZEUS results are quoted in this Table.}
      \label{leptonscombtable}
    \end{figure}

    ZEUS has analyzed 249~pb$^{-1}$ of data taken during the years 1998 to 2005 with a similar
    selection \cite{zeusiso2006}, and found the rate of such events at high hadronic transverse
    momentum to be consistent with the SM predictions (Fig.~\ref{leptonscombtable}). The excess
    observed by the H1 collaboration is not confirmed. Both experiments have studied differences
    in the acceptances and efficiencies of the respective analyses, and reached the conclusion
    that the experiments have comparable sensitivity in the regions where the H1 excess is
    observed \cite{h1iso2006,zeusiso2006}.

    Unfortunately, taking into account the amount of HERA data still to be analyzed in the future,
    it seems increasingly unlikely that the mystery of the isolated lepton events can be resolved
    by additional data alone.

  \subsection{High $\mathbf p_T$ Taus and Missing $\mathbf E_T$}

\begin{figure}[bth]
      \includegraphics[height=.25\textheight]{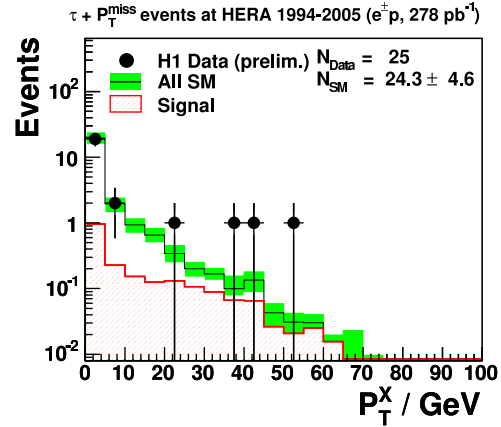}
      \caption{The hadronic transverse momentum distribution of $\tau + \MET$ events
	 in H1 data \cite{h1tau2006}. The SM expectation is shown as histogram with uncertainty
	 band. The signal component of the SM expectation, dominated by real $W$ production,
	 is given by the hatched histogram.}
      \label{h1tauscomb}
    \end{figure}

    In order to shed additional light on the question of isolated lepton production at HERA, both
    H1 and ZEUS have investigated the production of high $p_T$ tau leptons. In the latest H1
    analysis \cite{h1tau2006}, data corresponding to an integrated luminosity of 278~pb$^{-1}$
    have been used. The $\tau$ leptons are identified by using an identification algorithm based
    on the search for isolated charged tracks associated to narrow hadronic jets detected in the
    calorimeters, a typical signature of the one-prong hadronic $\tau$ decay. The 25 events found
    in the data are in good agreement with the SM expectation of $24.2^{+4.2}_{-5.8}$ events. In
    the region where the hadronic system has a transverse momentum $P_T^X > 25$~GeV, three events
    are observed in the data where the SM expectation is $0.74^{+0.19}_{-0.16}$ events. All three
    events have been collected in $e^-p$ collisions, in contrast to the excess observed in the
    $e$ and $\mu$ channels. The $p_T^X$ distribution is shown in Fig.~\ref{h1tauscomb}.

    The ZEUS collaboration has published the results of an analysis \cite{zeustau2004} of events
    containing isolated tau leptons and large missing transverse momentum based on 130~pb$^{-1}$ of
    HERA I data. For the $\tau$ identification, six observables based on the internal jet structure
    were exploited to discriminate between hadronic $\tau$ decays and quark- or gluon-induced jets.
    Three tau candidates were found, while $0.40^{+0.12}_{-0.13}$ were expected from SM processes.
    Requring $P_T^X > 25$~GeV, two candidate events remain, while $0.20 \pm 0.05$ events are expected
    from SM processes, about half of which is real $W$ production. Both events occured in $e^+p$
    collisions.

  \subsection{Events with Multiple Leptons}

\begin{figure}[bth]
      \includegraphics[height=.25\textheight,angle=270]{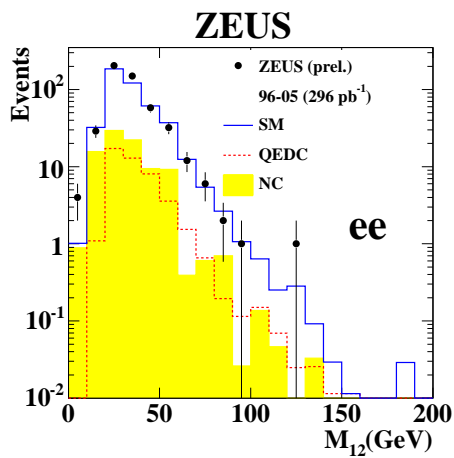}
      \caption{In the ZEUS analysis of events with two high $p_T$ electrons \cite{zeusmultilepton2006},
      comparison of the observed invariant mass $M_{12}$ of the two electrons with the SM
      expectation. The contributions of the QED Compton and neutral current DIS processes are also
      shown.}
      \label{zeus2e}
    \end{figure}

\begin{figure}[bth]
      \includegraphics[height=.25\textheight]{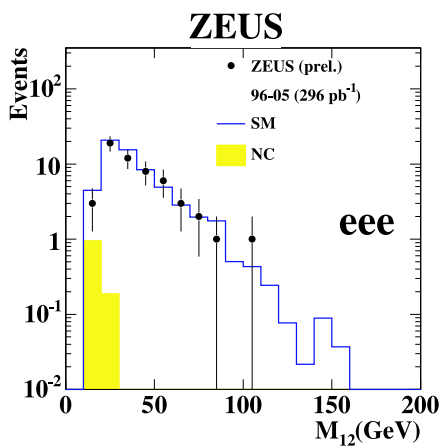}
      \caption{In the ZEUS analysis of events with three high $p_T$ electrons \cite{zeusmultilepton2006},
      comparison of the observed invariant mass $M_{12}$ of the two highest $p_T$ electrons with the SM
      expectation. The contributions of the QED Compton and neutral current DIS processes are also
      shown.}
      \label{zeus3e}
    \end{figure}

    Both ZEUS \cite{zeusmultilepton2006} and H1 \cite{h1multilepton2006} have studied the production of events
    containing multiple high $p_T$ isolated leptons. The dominant SM contribution to these final states is the two
    photon process $\gamma \gamma \rightarrow l^+ l^-$, which can be accurately predicted.
    ZEUS \cite{zeusmultilepton2006} has analyzed 296~pb$^{-1}$ of data for the $ee$ and $eee$ final states,
    and compared event yields and kinematical distributions, see for example Figs.~\ref{zeus2e} and
    \ref{zeus3e}. Data and SM expectations are found to be in agreement. The $\tau \tau$ final state has
    been analyzed in a smaller data set corresponding to 135~pb$^{-1}$, and again no deviation from
    the SM prediction has been found.

\begin{figure}[bth]
      \includegraphics[height=.32\textheight]{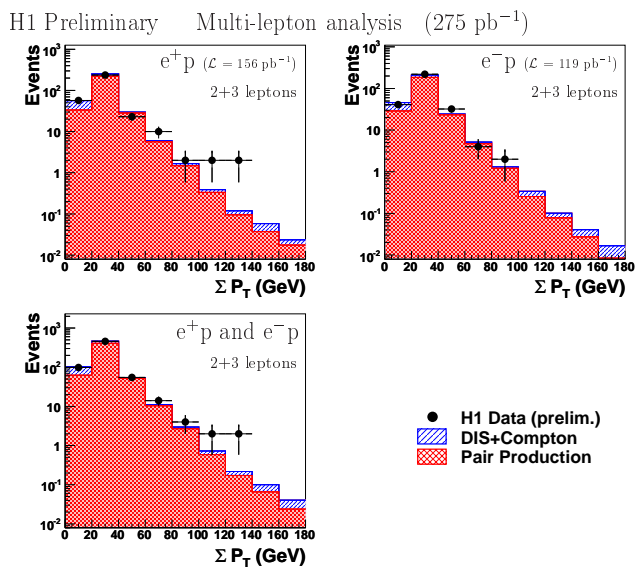}
      \caption{In the H1 analysis of events with two or three high $p_T$ electrons or
      muons \cite{h1multilepton2006}, distributions of the scalar sum of the lepton
      transverse momenta compared to the expectations, separately for $e^+p$ data,
      $e^-p$ data, and the entire data set.}
      \label{h1multileptons}
    \end{figure}

    The H1 collaboration has updated and extended their analysis of multi-lepton events \cite{h1multilepton2006},
    using 275~pb$^{-1}$ of data. The final states $ee$, $\mu\mu$, $e\mu$, $eee$ and $e\mu\mu$ have
    been studied for anomalies, see for example Fig.~\ref{h1multileptons} for the distributions of
    the scalar sum $\sum{p_T}$ of the lepton transverse momenta for all final states combined. For
    $\sum{p_T} > 100$~GeV, four events are observed with a SM expectation of $1.1 \pm 0.2$. All four
    events have been collected in $e^+p$ collisions, and three of the four events are in the $eee$
    final state and have an invariant mass of the two leading electrons of $M_{12}>100$~GeV.
    Apart from this moderate but interesting disagreement, data and SM expectation are found to be
    in good agreement.

  \subsection{Model Independent Search}

\begin{figure}[bth]
      \includegraphics[height=.22\textheight]{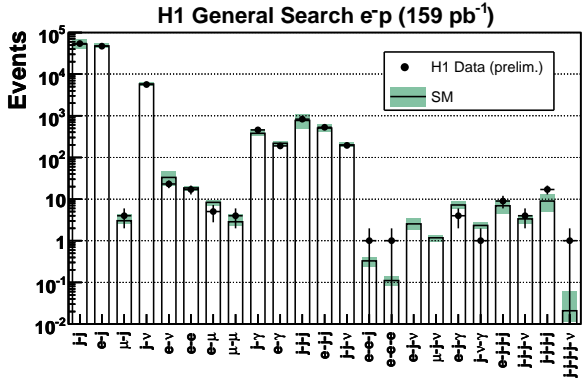}
      \caption{In the generic H1 search for deviations from the SM \cite{h1nomodel2006}, the
      observed number of events in the different exclusive event classes, as well as
      the SM expectation with its uncertainty. Only final states with either at least one
      data event or a SM expectation greater than one event are shown.}
      \label{h1nomodel}
    \end{figure}

    The H1 collaboration has previously presented an analysis based on the HERA I data using a model
    independent approach to search for deviations from the standard model, reporting no significant
    findings. Since the HERA I data consisted mostly of $e^+p$ data, the analysis has recently been
    updated with the $e^-p$ data collected in the years 2005--2006 and corresponding to an
    integrated luminosity of 159~pb$^{-1}$. Events are assigned to exclusive classes according to
    their final state involving isolated electrons, photons, muons, neutrinos (\MET) and jets with
    high transverse momenta. The event yields in the different classes are shown in
    Fig.~\ref{h1nomodel} together with the SM expectation. The $\mu-\nu$ class has been discarded
    because it is dominated by poorly reconstructed muons giving rise to large \MET. A statistical
    algorithm is applied to search for deviations from the SM in the distributions of the scalar sum
    of transverse momenta or the invariant mass of final state particles, and to quantify their
    significance. No significant deviation has been found.

\section{Supersymmetry}

  Supersymmetry or SUSY, a proposed invariance of nature for the  interchange of fermionic
  and bosonic degrees of freedom, has many important features which justify an intensive
  research at the highest energy accelerators. It allows for the unification of the four
  known forces, it is the only non-trivial extension of the Lorentz-Poincar\'e group, and it
  provides an elegant solution to evade the fine tuning problem of the standard model. In
  SUSY every SM particle has a partner differing in spin S by $1/2$. The SUSY partners are
  assigned an $R$-parity $R = (-1)^{3B+L+2S} = - 1$, where $B$ is the baryon and $L$ is the
  lepton number of the particle, in contrast to the SM particles of $R = +1$. A second Higgs
  doublet has to be introduced, leading to four additional Higgs particles which have $R =
  +1$. In the minimal supersymmetric extension of the SM, the MSSM, 105 additional
  parameters are introduced, corresponding to sparticle masses, mixing angles etc.

  Since the $R = -1$ partners have not yet been observed in nature, SUSY cannot be an exact
  symmetry. Various mechanisms for SUSY breaking have been proposed, each of them requiring a
  different set of new model parameters. Under certain assumptions the number of free
  parameters can be reduced to managable numbers. In the model that is probably most studied,
  minimal supergravity or mSUGRA, five parameters remain: the common scalar and fermion
  masses at the GUT scale, $m_0$ and $m_{1/2}$, the ratio of the vacuum expectation values of
  the two neutral Higgs fields $\tan\beta$, the trilinear coupling parameter $A_0$, and the
  sign of the higgsino mass parameter $\mu$. In the case of minimal gauge mediated SUSY
  breaking (mGMSB), the six parameters are $\Lambda$, $M_m$, $N_5$, $C_{grav}$, $\tan\beta$
  and the sign of $\mu$.

  In most cases $R$-parity is assumed to be conserved since there are severe limits on
  $B$ and $L$ violating processes. Then, the SUSY partners are pair produced and the
  lightest SUSY particle (LSP) is neutral and weakly interacting, and thus escapes
  detection. Therefore, the basic experimental signature for $R$-parity conserving SUSY is missing
  transverse energy and multiple jets and leptons originating from the cascade decay of
  the heavy $R = -1$ partners. Important SM backgrounds are $t\bar{t}$ production and
  gauge boson production either in pairs or accompanied by jets. In models with $R$-parity 
  conservation, the LSP, in most cases the \neutralino, is a natural dark matter candidate.

  Nevertheless, $R$-parity violation (\rpv) is not excluded, and remains an interesting
  alternative. \rpv\ would allow single resonant production of SUSY particles, and
  often even more jets or leptons in the final state from the $B$ or $L$ violating
  processes. In this case the theory contains 48 additional unknown Yukawa couplings.
  It is usually assumed that only one of these couplings is non-zero. At the Tevatron
  both $R$-parity conserving and \rpv\ processes have been studied.

  \subsection{Supersymmetry with $\mathbf R$-Parity Conserved}

  The searches for SUSY under the assumption of $R$-parity conservation are presented
  as follows: searches for charginos and neutralinos, and squarks and gluinos,
  in mSUGRA and mSUGRA inspired scenarios, which are benchmark processes at the
  Tevatron. Following this, results for alternative mechanisms of SUSY breaking are
  discussed, including split SUSY, GMSB, and anomaly mediated SUSY breaking.

  \subsubsection{Associated Production of Charginos and Neutralinos}

\begin{figure}[bth]
      \begin{tabular}{cc}
      \includegraphics[height=.11\textheight]{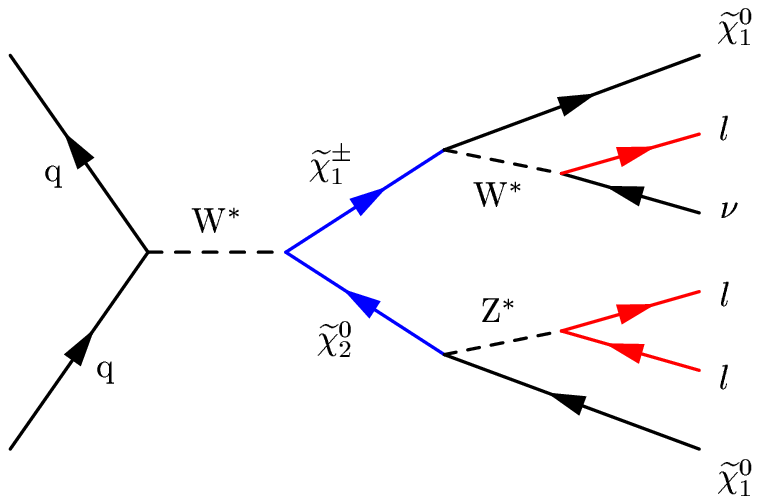} &
      \includegraphics[height=.11\textheight]{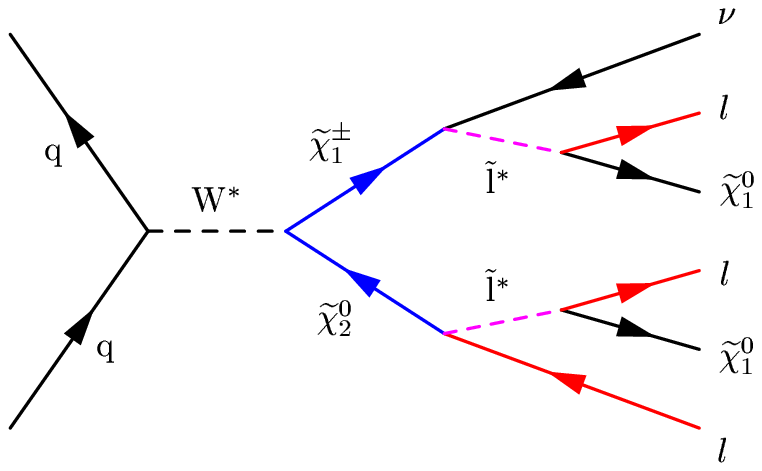}
      \end{tabular}
      \caption{Dominant production of charginos and neutralinos at the Tevatron and their
            decays.}
      \label{trileptondia}
    \end{figure}

\begin{figure}[bth]
      \includegraphics[height=.23\textheight]{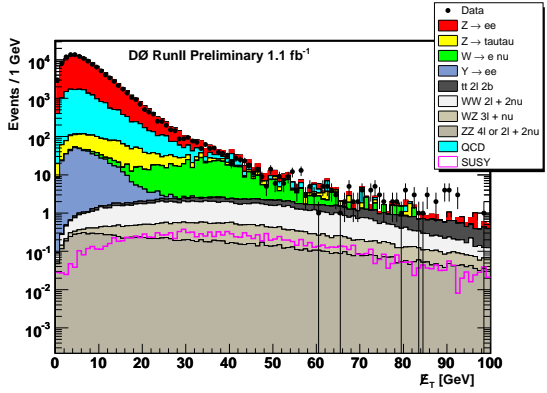}
      \caption{Distribution of \MET\ at the preselection stage of the ``eel'' chargino/neutralino search by D\O.}
      \label{trileptonmet}
    \end{figure}

    The dominant production mechanisms of charginos and neutralinos at the Tevatron along with
    their leptonic decays are depicted in Fig.~\ref{trileptondia}. The golden experimental
    signature searched for is three leptons, accompanied by \MET. For increased acceptance, the
    third lepton is sometimes identified as an isolated track, or not required to be found at all
    in the case of two leptons of same charge. The SM backgrounds ($Z/\gamma^* +$~jets, QCD
    (multijets), $WW/WZ$ and $t\bar{t}$ production) are small and well under control already at
    the preselection stage, as for example seen in Fig.~\ref{trileptonmet}.
    
\begin{figure}[bth]
      \includegraphics[height=.22\textheight]{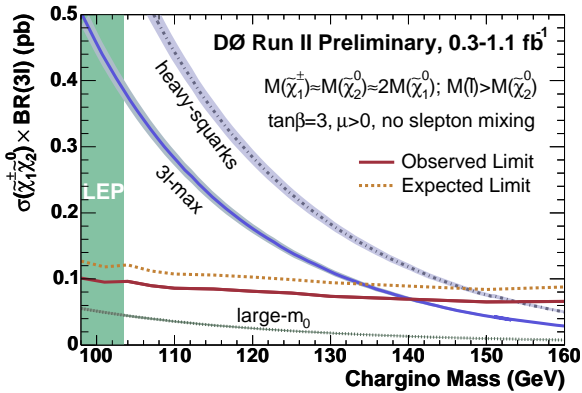}
      \caption{Cross section upper limits times branching fraction of chargino and
            neutralino production measured by D\O\ along with theoretical predictions.}
      \label{trilepton1}
    \end{figure}

\begin{figure}[bth]
      \includegraphics[height=.22\textheight]{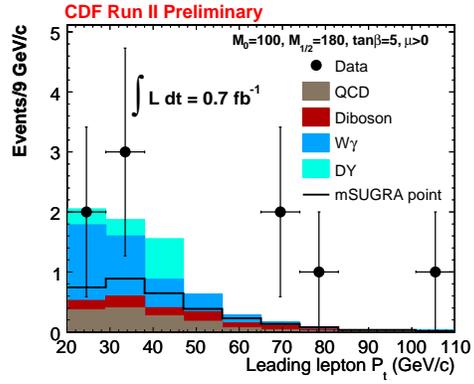}
      \caption{Transverse momentum of the leading lepton in the CDF search for charginos and neutralinos in
      like sign dileptons.}
      \label{cdfdilep}
    \end{figure}

    Because of the small leptonic branching fractions several final states need to be combined. In Table
    \ref{trilepton2} the results in twelve different channels studied by the two collaborations are
    listed. Since the observed number of events is in agreement in every channel with the predicted
    background, upper limits on the cross section times branching fraction were derived, as shown for
    example in the case of D\O\ in Fig.~\ref{trilepton1}. The theoretical predictions are for three mSUGRA
    inspired scenarios for mass relations as indicated in the figure. Lower limits of the chargino mass
    have been derived for two scenarios with large leptonic branching fractions: $m(\chargino) < 140$~GeV
    is excluded when the slepton mass is slightly above the mass of the second neutralino, thus allowing
    only 3-body decays (denoted ``$3l$-max''), and $m(\chargino) < 154$~GeV is excluded for the case when
    squarks are heavy and therefore the destructive $t$-channel contribution is minimal. The CDF analyses
    have comparable sensitivity. The slight excess in the CDF same sign dilepton channel is worth noting,
    because it turns out that four of the nine events have a leading lepton with high transverse momentum
    in excess of 60~GeV, where neither SM background nor the SUSY signal are expected
    (Fig.~\ref{cdfdilep}).

    \begin{table}
    \begin{tabular}{c|c|c|c}
      D\O\ channels       & $\cal{L}_{\mbox{int}}$ [fb$^{-1}$] &   Background    & Data \\[1mm]\hline
      $ee + \mbox{track}$                                & 1.1\ & $0.76 \pm 0.67$ & 0 \\
      $\mu\mu + \mbox{track}$                            & 0.32 & $1.75 \pm 0.57$ & 2 \\
      $e\mu + \mbox{track}$                              & 0.32 & $0.31 \pm 0.13$ & 0 \\
      $\mu^+\mu^+ / \mu^-\mu^-$                          & 0.9\ & $1.1  \pm 0.4 $ & 1 \\
      $e\tau + \mbox{track}$                             & 0.33 & $0.58 \pm 0.14$ & 0 \\
      $\mu\tau + \mbox{track}$                           & 0.33 & $0.36 \pm 0.13$ & 1 \\\hline\hline
      CDF channels  & $\cal{L}_{      \mbox{int}}$ [fb$^{-1}$] &   Background    & Data \\[1mm]\hline
      $ee + e/\mu$                                       & 0.35 & $0.17 \pm 0.05$ & 0 \\
      $ee + \mbox{track}$                                & 0.61 & $0.49 \pm 0.10$ & 1 \\
      $\mu\mu + e/\mu$ (low $p_T$)                       & 0.31 & $0.13 \pm 0.03$ & 0 \\
      $\mu\mu + e/\mu$ (high $p_T$)                      & 0.75 & $0.64 \pm 0.18$ & 1 \\
      $e\mu + e/\mu$                                     & 0.75 & $0.78 \pm 0.11$ & 0 \\
      $e^\pm e^\pm$, $e^\pm \mu^\pm$, $\mu^\pm \mu^\pm$  & 0.70 & $6.8  \pm 1.0$  & 9 \\
    \end{tabular}
    \caption{Three and two lepton final states studied by the D\O\ and CDF collaborations in the
    search for charginos and neutralinos.}
    \label{trilepton2}
    \end{table}

  \subsubsection{Squarks and Gluinos}

\begin{figure}[bth]
      \includegraphics[height=.2\textheight]{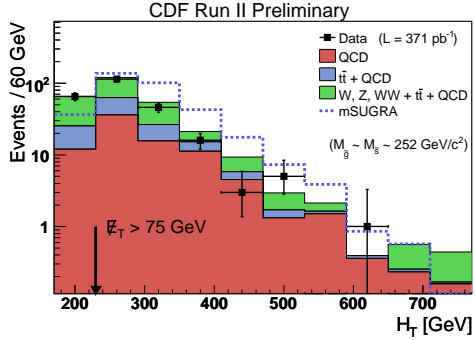}
      \caption{Distribution of $H_T$ obtained by CDF for the 3-jet event topology and a light
       gluino, compared with the SM background and the expected SUSY signal.}
      \label{sqgl1}
    \end{figure}

\begin{figure}[bth]
      \includegraphics[height=.28\textheight]{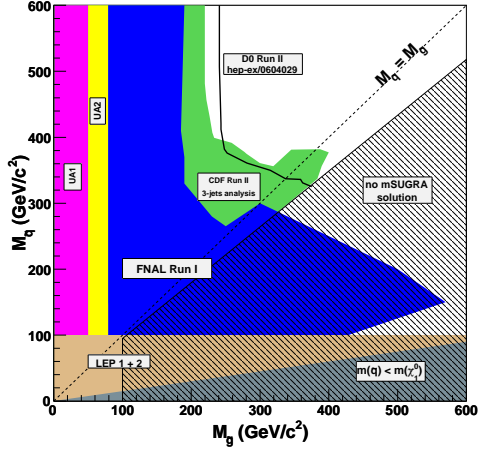}
      \caption{Excluded regions in the $(M_{\tilde{q}}$, $M_{\tilde{g}})$ plane obtained by CDF and D\O.
	It should be noted that D\O\ has used conservatively a theoretical cross section reduced by its
	uncertainty when calculating limits.}
      \label{sqgl2}
    \end{figure}

    D\O\ has carried out a generic search \cite{sqglpub} for gluinos and squarks requiring a minimum
    number of jets, $N_j$, accompanied by substantial \MET\ and $H_T$, the scalar sum of the jet
    transverse energies, for the following three topologies: (i) $N_j = 2$ for $M_{\tilde{q}} <
    M_{\tilde{g}}$, (ii) $N_j = 3$ for $M_{\tilde{q}} \simeq M_{\tilde{g}}$, and (iii) $N_j = 4$ for
    $M_{\tilde{q}} > M_{\tilde{g}}$, where $M_{\tilde{q}}$ and $M_{\tilde{g}}$ are the mass of the
    squark and the mass of the gluino, respectively, and the jet multiplicities are chosen corresponding
    to the decay modes $\tilde{q} \rightarrow q \neutralino$ and $\tilde{g} \rightarrow q \bar{q}
    \neutralino$. A general agreement between the data corresponding to an integrated luminosity of
    310~pb$^{-1}$ and the expected background at all stages of the selection is observed, for the
    D\O\ analyses as well as for the CDF analysis in the 3-jet final state, as can be seen
    for example in Fig.~\ref{sqgl1}, where the CDF $H_T$ distribution optimized to search for relatively
    small gluino masses is displayed. The three
    D\O\ analyses optimized for the three different mass hierarchies are combined, and the exclusion region
    in the $(M_{\tilde{q}}$, $M_{\tilde{g}})$ plane as shown in Fig.~\ref{sqgl2} is obtained. Also shown
    is the excluded region obtained by CDF using the 3-jet event topology and 371~pb$^{-1}$ of data. The
    absolute lower mass limits at 95\% C.L. are $M_{\tilde{q}} > 325$~GeV and $M_{\tilde{g}} > 241$~GeV,
    while for $M_{\tilde{q}} \simeq M_{\tilde{g}}$, $M_{\tilde{q}, \tilde{g}} < 387$~GeV can be excluded.

    Third generation squarks may be light due to the large mixing between the scalar partners of
    the left and right handed quarks. They may be accessible at the Tevatron and are therefore
    subject of dedicated analyses by the two collaborations.

\begin{figure}[bth]
      \includegraphics[height=.2\textheight]{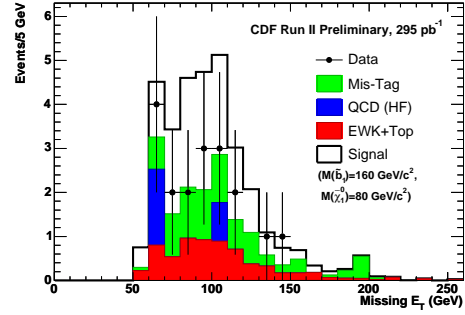}
      \caption{Distribution of \MET\ in events with at least one $b$-tagged jet observed by CDF
        in the search for sbottom pair production with $\tilde{b} \rightarrow b \neutralino$, together
	with the SM background and the expected signal.}
      \label{sbottom1}
    \end{figure}

\begin{figure}[bth]
      \includegraphics[height=.27\textheight]{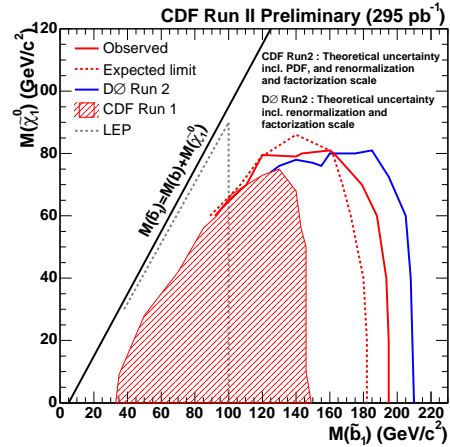}
      \caption{Mass values of the sbottom and neutralino excluded by the D\O\ and CDF analyses of
        sbottom pair production followed by the decay $\tilde{b} \rightarrow b \neutralino$.}
      \label{sbottom2}
    \end{figure}

    Both D\O\ \cite{sbpub} ($\cal{L}_{\mbox{int}} = $ 310~pb$^{-1}$) and CDF (295~pb$^{-1}$) have
    searched for direct pair production of the lightest sbottom quark $\tilde{b}_1$, assuming that it
    decays with a branching fraction of 100\% into a $b$ quark and the lightest neutralino. The
    experimental signature is two acoplanar $b$ jets and \MET. In both analyses at least one of the
    jets was required to be identified as a $b$ jet using lifetime information. The selection value of
    the \MET\ and that of the $E_T$ of the jets were optimized according to the mass value of the
    sbottom to be detected. The data did not show any significant excess over the expected SM
    background as can be seen for example in Fig.~\ref{sbottom1}, where the \MET\ distribution
    obtained by CDF is shown. The excluded mass values of the sbottom and the neutralino are shown in
    Fig.~\ref{sbottom2}, substantially improving on previous limits.

\begin{figure}[bth]
      \includegraphics[height=.2\textheight]{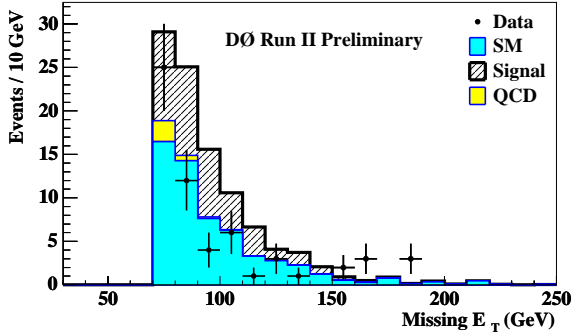}
      \caption{Final \MET\ distribution in events with at least one $c$-tagged jet observed by D\O\ in
        the search for stop pair production with $\tilde{t} \rightarrow c \neutralino$,
	together with the SM background and the expected signal for $M_{\tilde{t}_1} = 130$~GeV and
	$M_{\neutralino} = 50$~GeV.}
      \label{stop1}
    \end{figure}

\begin{figure}[bth]
      \includegraphics[height=.27\textheight]{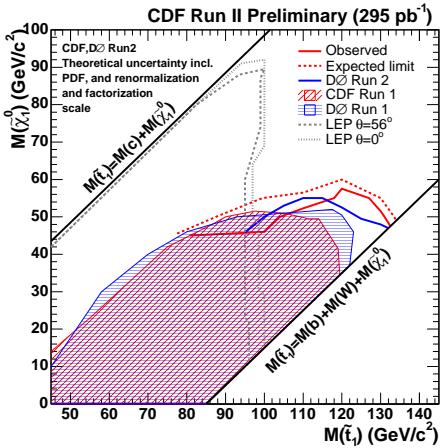}
      \caption{Mass values of the lightest stop and neutralino excluded by the D\O\ and CDF analyses of
         stop pair production followed by the decay $\tilde{t} \rightarrow c \neutralino$.}
      \label{stop2}
    \end{figure}

    If the mass $M_{\tilde{t}_1}$ of the lightest stop quark $\tilde{t}_1$ satisfies the
    relation $M_c + M_{\neutralino} < M_{\tilde{t}_1} < M_b + M_W + M_{\neutralino}$,
    where $M_c$, $M_b$, $M_{\neutralino}$ and $M_W$ are the masses of the $c$ quark, the
    $b$ quark, the lightest neutralino and the $W$ boson, respectively, its dominant decay
    mode is $\tilde{t}_1 \rightarrow c \neutralino$, and a similar search strategy can be
    applied as for the sbottom analysis outlined above, except that jets should satisfy a
    $c$-tag instead of a $b$-tag criterion. Again, data observed by D\O\ and by CDF are in
    agreement with the SM expectation (see for example Fig.~\ref{stop1}), and exclusion
    limits can be set in the mass plane of the lightest stop and the neutralino, extending
    the previously excluded regions of $M_{\tilde{t}_1}$ and $M_{\neutralino}$, as shown
    in Fig.~\ref{stop2}.

\begin{figure}[bth]
      \includegraphics[height=.3\textheight]{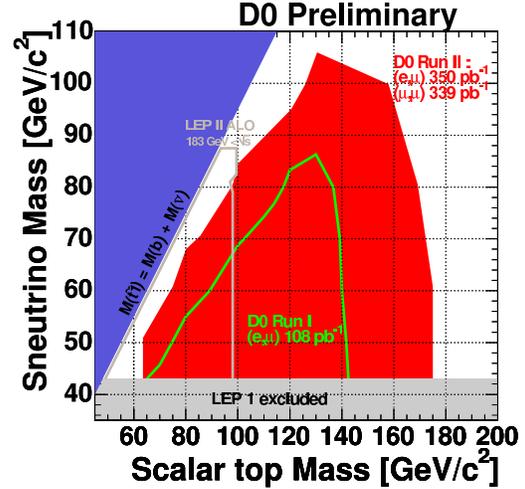}
      \caption{Excluded regions of the stop and sneutrino masses obtained by D\O\ assuming
	 the decay $\tilde{t}_1 \rightarrow b + l + \tilde{\nu}$ ($l = e$, $\mu$).}
      \label{blsnu}
    \end{figure}

    D\O\ has also searched for pair production of $\tilde{t}_1$ quarks assuming that they decay
    into a $b$ quark, a lepton and a sneutrino via a virtual chargino, which may be favorable
    due to the relatively weak constraint on the sneutrino mass from LEP, $M_{\tilde{\nu}} >
    43.7$~GeV. The final state is two isolated leptons with opposite charge, two $b$ jets, and
    \MET. For the two leptons, the combinations $\mu\mu$ and $e\mu$ have been analyzed. While
    in the $\mu\mu$ analysis a $b$ jet is required to be identified, the $e\mu$ analysis has
    smaller backgrounds and requires a certain number of non-isolated tracks instead of an
    explicit jet reconstruction, thus increasing the sensitivity. In neither channel a
    significant signal for the presence of the $\tilde{t}_1$ quark has been observed.
    Therefore, the two analyses have been combined to exclude masses of the $\tilde{t}_1$ quark
    and the sneutrino. The mass region excluded by earlier experiments has been significantly
    extended, as can be seen in Fig.~\ref{blsnu}.

  \subsubsection{Split SUSY}

    Split SUSY is a relatively new variant of supersymmetry in which the supersymmetric scalars
    are heavy (possibly GUT scale) compared to the (SUSY) fermions \cite{split}. It avoids much of
    the fine tuning required to remain consistent with observations while still preserving the
    favored consequences: a dark-matter candidate is still present in the theory to explain the
    observed cold dark matter density in the universe, and coupling unification at the GUT scale
    still occurs. Due to the high masses of the scalars, the gluino decays are suppressed. The
    gluinos have time to hadronize into ``R-hadrons'', colorless bound states of a gluino and
    other quarks or gluons. At the Tevatron, R-hadrons could be pair produced copiously through
    strong interactions. About half of the R-hadrons are expected to be charged, and some charged
    R-hadrons can become ``stopped gluinos'' by losing all of their momentum through ionization
    and coming to rest in the calorimeters. Their decays may happen after several bunch crossings.

\begin{figure}[bth]
      \includegraphics[height=.22\textheight]{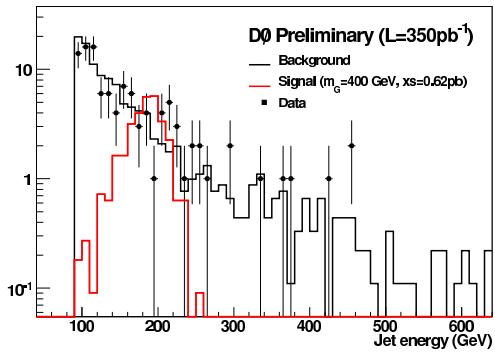}
      \caption{In the D\O\ search for stopped gluinos, energy distribution of the selected
       jets (dots with error bars), of the
       background (black histogram) -- dominated by cosmic muon events --, and of a gluino of
       mass 400~GeV decaying into a
       gluon and a neutralino of 90~GeV mass.}
      \label{gluino1}
    \end{figure}

\begin{figure}[bth]
      \includegraphics[height=.22\textheight]{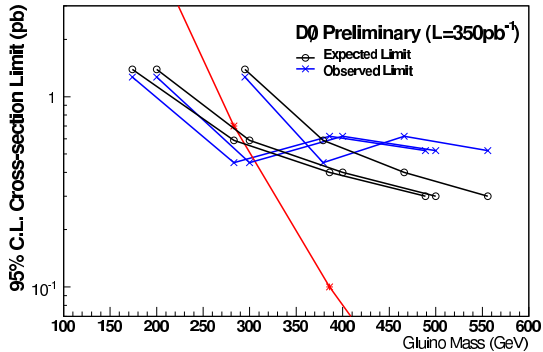}
      \caption{Expected and observed cross section upper limits from D\O\ for stopped gluinos
      as a function of the gluino mass for neutralino masses of 50~GeV,
      90~GeV and 200~GeV. Also shown is the theoretical cross section (red star).}
      \label{gluino2}
    \end{figure}

    D\O\ has searched for stopped gluinos in the $\tilde{g} \rightarrow g \neutralino$ decay channel by
    looking for randomly oriented jets in bunch crossings without an inelastic $p\bar{p}$
    collision. The background consists mainly of cosmic and beam halo muon induced jets where the
    muon escapes reconstruction. It has been estimated using the data. The energy distribution of
    the observed jets is shown in Fig.~\ref{gluino1}, along with that of the estimated background.
    As can be seen the data do not show any excess above the expected background. The derived cross
    section upper limits are shown in Fig.~\ref{gluino2} as a function of the gluino mass for
    different mass values of the $\neutralino$. The theoretical cross section is also shown, from which
    gluino masses below $\simeq 300$~GeV can be excluded if the mass of the $\neutralino$ is less than
    90~GeV. It is worth noting that the signature of a long-lived gluino could occur in many
    models of beyond the standard model physics.

  \subsubsection{Gauge Mediated SUSY Breaking}

\begin{figure}[bth]
      \includegraphics[height=.2\textheight]{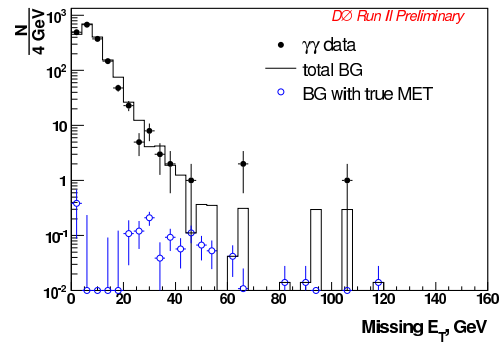}
      \caption{In events with two photons, \MET\ distribution of recent D\O\ data (full circles), of
	 the background (histogram), and the fraction of the latter with true \MET\ (open circles).}
      \label{gmsb1}
    \end{figure}

\begin{figure}[bth]
      \includegraphics[height=.22\textheight]{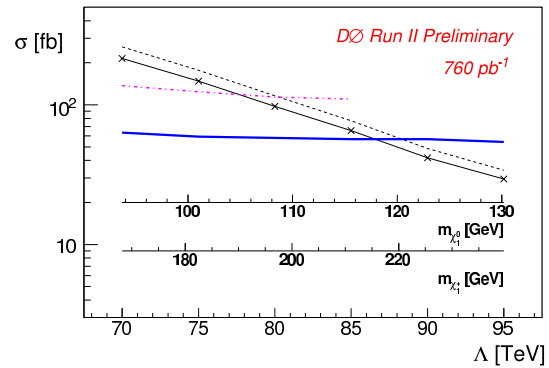}
      \caption{Cross section upper limits (solid blue line) in the D\O\ GMSB analysis with
      two photons and \MET. The theoretical LO (NLO) cross section for the GMSB process
      (solid (dashed) black lines) corresponds to the SUSY Snowmass slope 8, with a
      messenger mass $M_m = 2\Lambda$, the number of messengers $N_5=1$, $\tan\beta = 15$,
      and $\mu > 0$.}
      \label{gmsb2}
    \end{figure}

    In the mGMSB scenario, the gravitino is the LSP with a mass less than $\sim 1$~keV, and the
    next-to-lightest SUSY particle (NLSP) may be the lightest neutralino, which decays to a
    gravitino and a photon. The lifetime of the NLSP is a priori unknown and depends on the
    $C_{grav}$ parameter of the model. D\O\ and CDF have searched for two highly energetic
    prompt photons accompanied by large \MET, caused by the undetected gravitino. As can be seen
    in Fig.~\ref{gmsb1}, the observed \MET\ distribution is compatible with the SM background.
    With recent data corresponding to 760~pb$^{-1}$, D\O\ has improved the previous limits
    obtained from combined measurements of CDF and D\O, see Fig.~\ref{gmsb2}. For the $\Lambda$
    parameter, determining the effective scale of SUSY breaking, $\Lambda < 88.5$~TeV is
    excluded at 95\% C.L. This corresponds to a $\neutralino$ mass of $> 120$~GeV and a
    $\chargino$ mass of $> 220$~GeV.

\begin{figure}[bth]
      \includegraphics[height=.22\textheight]{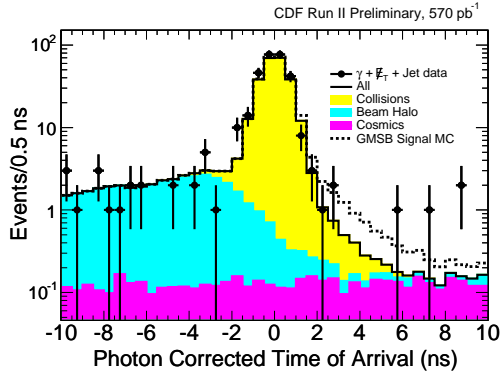}
      \caption{Photon arrival time distribution in the CDF electromagnetic calorimeter. The
      expected GMSB signal is for a neutralino mass of $93.6$~GeV with a lifetime of 10~ns.}
      \label{emtiming1}
    \end{figure}

\begin{figure}[bth]
      \includegraphics[height=.26\textheight]{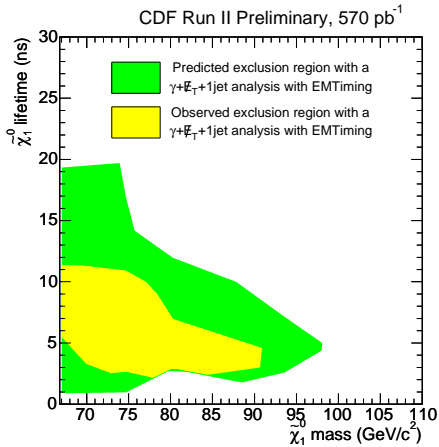}
      \caption{Excluded mass and lifetime values of the lightest neutralino (NLSP) in
        the CDF GMSB analysis with late photons.}
      \label{emtiming2}
    \end{figure}

    CDF searched for signs of GMSB assuming long lifetime for the NLSP, i.e.~the
    neutralino, by looking for ``late'' photons in the calorimeter, using 570~pb$^{-1}$
    of data. Fig.~\ref{emtiming1} shows that the arrival time distribution of photons
    accompanied by \MET\ doesn't show a significant excess at positive times where the
    GMSB signal is expected. Ten events are observed with $7.6 \pm 1.9$ background
    events expected. This allows to exclude a region in the plane of the neutralino
    lifetime versus the neutralino mass, as shown in Fig.~\ref{emtiming2}.

  \subsubsection{Anomaly Mediated SUSY Breaking}

\begin{figure}[bth]
      \includegraphics[height=.23\textheight]{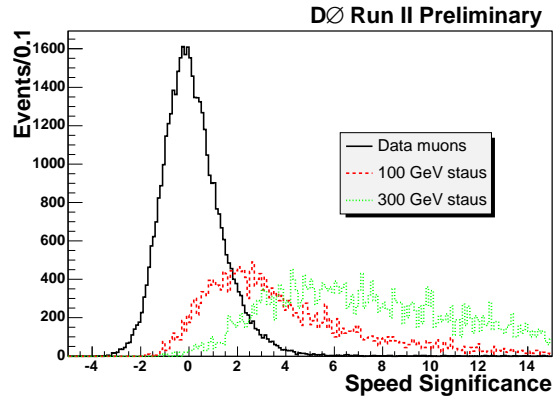}
      \caption{Speed significance distributions of muons and of long-lived staus
        for two different masses. The distributions are expected to be similar for charginos
	of the same mass.}
      \label{champs1}
    \end{figure}

\begin{figure}[bth]
     \includegraphics[height=.24\textheight]{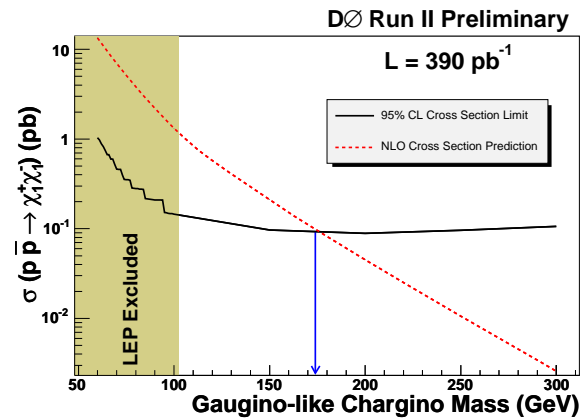}
      \caption{Cross section upper limit obtained by D\O\ and
       theoretical cross section for a gaugino-like chargino as a function of its mass.}
      \label{champs2}
    \end{figure}

    Many SUSY scenarios contain charged massive long-lived particles expected to traverse
    the entire detector, for example staus or charginos with long lifetime. D\O\ has
    studied chargino pair production assuming anomaly mediated supersymmetry breaking. In
    this case the mass difference of the lightest chargino and neutralino is expected to
    be small, less than $\simeq 150$~MeV. Therefore charginos can have a long lifetime,
    and leave a muon-like signature in the detector. However, due to their mass they move
    slowly, and the speed significance, defined as $s = (1-v)/\sigma_v$, where $v$ is the
    speed of the chargino in units of the speed of light as measured in the muon system
    and $\sigma_v$ is its uncertainty, is shifted towards positive values depending on the mass,
    as indicated in Fig.~\ref{champs1} for heavy staus. In the D\O\ analysis of
    390~pb$^{-1}$ of data, two muons with $s > 0$ were required, and a final optimized cut
    was placed in the plane of the dimuon invariant mass and $s$. The observed number of
    events is compatible with the expected background, estimated from muon pairs from the
    $Z \rightarrow \mu^+\mu^-$ decay. An upper limit of the production cross section was
    derived (Fig.~\ref{champs2}) which, by confronting it with the theoretical cross section,
    excludes wino-like charginos with a mass of less than 174~GeV.

\subsection{$\mathbf R$-Parity Violation}

  While in many analyses $R$-parity is assumed to be conserved, which leaves the lightest
  supersymmetric particle (LSP) stable, SUSY does not require $R$-parity conservation. If
  \rpv\ is allowed, the following trilinear and bilinear terms appear in the
  superpotential \cite{rpv}:
  \begin{eqnarray*}
  W_{\rpv}&=& + \frac{1}{2} \lambda _{ijk} L_i L_j \bar{E}_k
              + \lambda'_{ijk} L_i Q_j \bar{D}_k \\
          & & + \frac{1}{2} \lambda'' _{ijk} \bar{U}_i \bar{D}_j \bar{D}_k
              + \mu_i L_i H_u
  \end{eqnarray*}
  where $L$ and $Q$ are the lepton and quark SU(2) doublet superfields, $\bar{E}$,
  $\bar{U}$, $\bar{D}$ denote the singlet fields, and $i,j,k$ refer to the fermion families.
  The first two terms imply lepton number violation, while the third term leads to baryon
  number violation. The coupling strengths are given by the Yukawa coupling constants
  $\lambda, \lambda'$ and $\lambda''$. The last term, $\mu_i L_i H_u$, mixes the lepton and
  the Higgs superfields. The $\lambda$ and $\lambda'$ couplings give rise to final states
  with multiple leptons, which provide excellent signatures at the Tevatron. Stringent
  limits exist on the size of many \rpv\ couplings \cite{referbound,barbier}, in particular
  for the case of more than one non-zero coupling.

  \subsubsection{Gaugino Production in Multi Lepton Final States}

\begin{figure}[b]
    \includegraphics[height=.17\textheight]{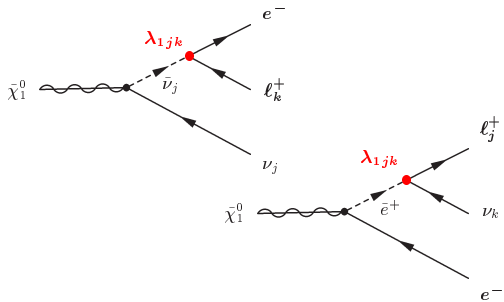}
    \caption{Two examples of \rpv-decays of the lightest neutralino via the \lle\ couplings
    $\lambda_{1jk}$. In each decay, two charged leptons and one neutrino are produced.}
    \label{rpvdecay}
  \end{figure}

  The charginos and neutralinos are produced in pairs or associated and with $R$-parity
  conserved. The produced sparticles (cascade) decay to the lightest neutralino
  \neutralino. Under the assumption of a single non-zero \lle\ coupling, the neutralino
  decays into two charged leptons and one neutrino by violating $R$-parity
  (Fig.~\ref{rpvdecay}). The final state therefore contains at least four charged leptons
  and two neutrinos which lead to missing transverse energy in the detector.

  Both D\O\ and CDF have searched for this signature, assuming a sufficiently large \lle\
  coupling leading to a prompt decay of the lightest neutrlino, i.e. the corresponding
  \lle-coupling ($\lambda_{121}$, $\lambda_{122}$, or $\lambda_{133}$) has to be larger than
  $\gtrsim 0.01$. In the D\O\ analysis \cite{lle}, based on 360~pb$^{-1}$ of data, for best
  acceptance only three charged leptons are required to be identified. Three different analyses
  $eel$, $\mu\mu l$ and $ee\tau$ with $l=e,\mu$ are performed depending on the flavors of the
  leptons in the final state. All three analyses are optimized separately using SM and signal
  Monte Carlo simulations. After all cuts, no events remain in the data, while $0.9 \pm 0.4$,
  $0.4 \pm 0.1$, and $1.3 \pm 1.8$ events are expected from SM processes in the $eel$, $\mu\mu
  l$, and $ee\tau$ analysis, respectively. The dominant SM backgrounds are due to $Z/\gamma^*
  \rightarrow l^+l^-$ and diboson production.

\begin{figure}[b]
    \includegraphics[height=.26\textheight]{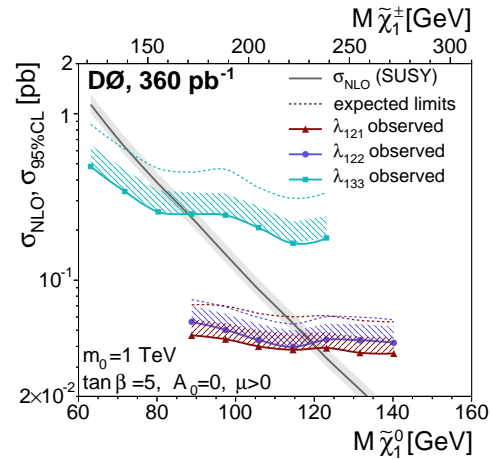}
    \caption{Cross section limits from D\O\ for three different \lle\ couplings compared to
	     the mSUGRA cross section prediction.}
    \label{lle1}
  \end{figure}

\begin{figure}[b]
    \includegraphics[height=.26\textheight]{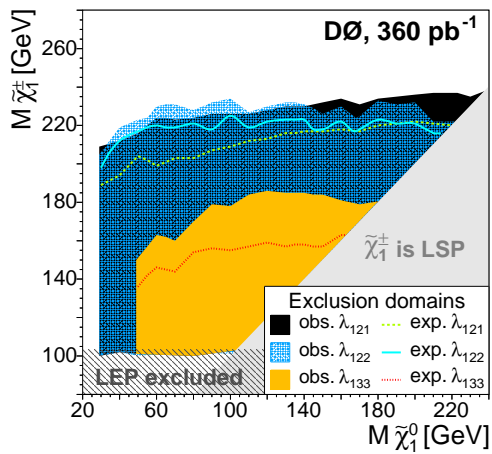}
    \caption{Exclusion contours for the three \lle\ couplings
	     $\lambda_{121}$, $\lambda_{122}$, $\lambda_{133}$ in the
	     m($\tilde{\chi}^0_1$)--m($\tilde{\chi}^\pm_1$) plane for an MSSM
             scenario without mass unification of $M_1$ and $M_2$.}
    \label{lle2}
  \end{figure}

  Since no evidence for \rpv-SUSY is observed, the analyses are combined and upper limits on
  the chargino and neutralino pair production cross section are set. Lower bounds on the
  masses of the lightest neutralino and the lightest chargino are derived in mSUGRA and in an
  MSSM scenario with heavy sfermions, but assuming no GUT relation between $M_1$ and $M_2$. The
  limits as shown in Figs.~\ref{lle1} and \ref{lle2} are the most restrictive to date. CDF has
  shown preliminary results with comparable sensitivity.

  \subsubsection{Long-Lived Neutral Particles in Dimuon Final States}

  In this D\O\ analysis \cite{nllp}, a small \lle\ coupling $\lambda_{122}$ is
  assumed, leading to a long neutralino $\tilde{\chi}^0_1$ lifetime and consequently to a
  displaced dimuon vertex. The final state is of particular interest due to an anomaly reported by
  the NuTeV collaboration; in 2000, NuTeV reported \cite{nutev} on a search for heavy neutral
  leptons decaying to $\mu\mu\nu$, amongst other final states. In the dimuon channel, they
  observed three events while only 0.07 events were expected. Because of the asymmetry in the muon
  momenta, and the absence of a signal in other channels, NuTeV argued that the signal was unlikely
  to be due to neutral heavy leptons.

\begin{figure}[bth]
    \includegraphics[height=.2\textheight]{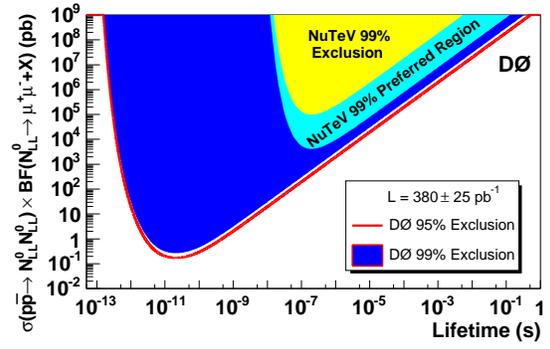} 
    \caption{Cross section upper limits times branching fraction for neutral
             long-lived particles decaying to muon pairs as a function of their lifetime.}
    \label{nllpfig}
  \end{figure}

  To shed light on the origin of these events, D\O\ has searched for pairs of oppositely charged
  isolated muons originating from a common vertex located between 5 and 20~cm radially displaced
  from the beamline. In this region, a good calibration using $K_S$ mesons is possible. No events
  have been found in 380~pb$^{-1}$ of data, with an estimated background of $0.8 \pm 1.6$ events.
  The result is interpreted as cross section upper limit on the production times branching
  fraction of a neutral long-lived particle decaying into $\mu^+\mu^- + X$ as a function of its
  lifetime, as shown in Fig.~\ref{nllpfig}. To compare with the NuTeV result, the momentum of the
  hypothetical new particles in the neutrino beam was converted to the Tevatron center of mass
  frame. While the result is somewhat dependent on the assumptions made regarding the decay of the
  long-lived particle, this result excludes an interpretation of the NuTeV excess of dimuon events
  in a large class of models.

  \subsubsection{Resonant Second Generation Slepton Production}

  The \lqd\ coupling offers the opportunity to produce sleptons in $p\bar{p}$ collisions as
  resonances. For a non-zero coupling $\lambda'_{211}$ this is either a smuon or a muon
  sneutrino. The slepton cascade decays into the lightest neutralino \neutralino\ and associated
  leptons. The neutralino decays via the same $R$-parity violating coupling $\lambda'_{211}$
  into a $2^{nd}$ generation lepton and two jets. The cross section is proportional to
  $(\lambda'_{211})^2$, so that limits on this coupling can be derived.

  D\O\ has recently published \cite{lqd} the results of a search for resonant second
  generation slepton production. The three channels (i)
  $\tilde{\mu}\to\tilde{\chi}^0_1\,\mu$, (ii) $\tilde{\mu}\to\tilde{\chi}^0_{2,3,4}\,\mu$,
  and (iii) $\tilde{\nu}_\mu\to\tilde{\chi}^\pm_{1,2}\,\mu$ resulting in dimuon and
  multijet final states are analyzed separately. For the further discrimination of the
  signal and the SM background, the analysis makes use of the possibility to reconstruct
  the neutralino and the slepton masses: using the leading muon and the two leading jets
  one would be able to reconstruct the lightest neutralino, and a peak in the invariant
  mass of the two muons and all jets would indicate the presence of the slepton. The
  selection criteria are optimized depending on the slepton and neutralino mass, and
  for all 117 mass combinations being probed, the data corresponding to 380~pb$^{-1}$
  show no excess with respect to the SM expectation.

\begin{figure}[bth] \setlength{\unitlength}{1cm}
    \begin{picture}(8.6,6.8)(0.0,0.0)
      \put(0.8,0.2) {\includegraphics[height=.26\textheight]{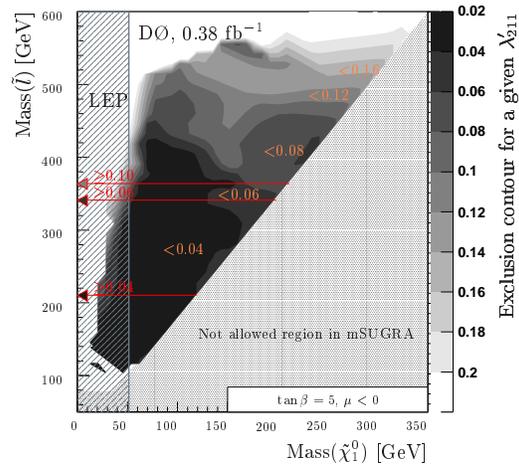}}
    \end{picture}
    \caption{Cross section upper limits for the slepton mass versus the neutralino
             mass obtained in the D\O\ search for \rpv\ SUSY with a non-zero
	     $\lambda'_{211}$ coupling.}
    \label{slep}
  \end{figure}

  In the absence of an excess in the data, cross section limits on resonant slepton
  production were set. The results are interpreted within the mSUGRA framework with
  $\tan\beta=5$, $\mu<0$ and $A_0=0$, and an exclusion contour as a function of
  $\lambda'_{211}$ is derived after combination of all three channels, as shown in
  Fig.~\ref{slep}. Lower limits for the slepton mass of 210, 340 and 363~GeV are
  obtained for $\lambda'_{211}$ values of 0.04, 0.06 and 0.10, respectively, a
  significant improvement compared to previous results.

  \subsubsection{Resonant Sneutrino Production}

\begin{figure}[bth]
    \includegraphics[height=.15\textheight]{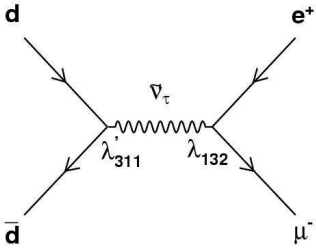}
    \caption{Production and decay of a tau sneutrino $\tilde{\nu}_\tau$ involving
      two different \rpv\ couplings.}
    \label{rpvsnu1}
  \end{figure}

\begin{figure}[bth]
    \includegraphics[height=.25\textheight]{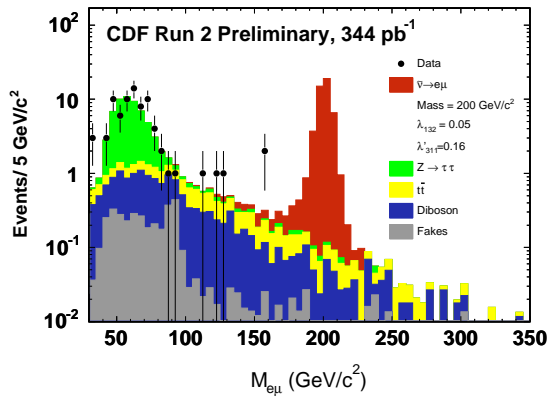}
    \caption{Invariant mass distribution of the electron-muon pairs in the CDF search for
    resonant tau sneutrino production.}
    \label{rpvsnu2}
  \end{figure}

\begin{figure}[bth]
    \includegraphics[height=.25\textheight]{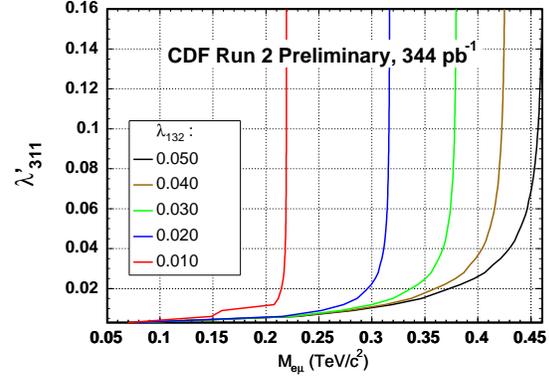}
    \caption{Excluded regions in the plane $\lambda'_{311}$
        versus the mass of the $\tilde{\nu}_\tau$, obtained by CDF.}
    \label{rpvsnu3}
  \end{figure}

  Here, a $\tilde{\nu}_\tau$ is produced with a $\lambda'_{311}$ Yukawa coupling which
  subsequently decays to an oppositely charged electron-muon pair via a non-zero
  $\lambda_{132}$ coupling (Fig.~\ref{rpvsnu1}). If such a process existed, a peak in the
  electron-muon invariant mass would be seen. CDF searched for such electron-muon pairs in
  344~pb$^{-1}$ of data, but has seen no indication for an enhancement in their mass
  distribution (Fig.~\ref{rpvsnu2}). Therefore exclusion limits as a function of the
  $\tilde{\nu}_\tau$ mass and the two Yukawa couplings are derived as shown for example in
  Fig.~\ref{rpvsnu3}.

  \subsubsection{Stop Pair Production}

\begin{figure}[bth]
    \includegraphics[height=.2\textheight]{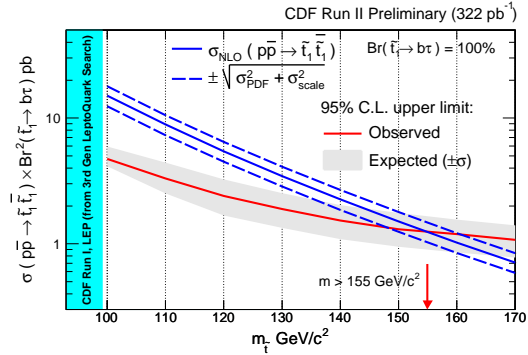}
    \caption{Upper limits of cross section times
       branching fraction as a function of the stop mass obtained by CDF assuming the
       decay $\tilde{t}_1 \rightarrow b + \tau$. Also shown is the theoretical expectation.}
    \label{rpvstop}
  \end{figure}

  CDF has searched for pair production of stop quarks where both of them decay promptly into a $b$ quark
  and a $\tau$ via the $R$-parity violating non-zero $\lambda '_{333}$ coupling, and one of the taus
  decays leptonically, the other one into hadrons. In 322~pb$^{-1}$ of data, two events have been
  found with an isolated electron (or muon), a hadronic tau decay and two additional jets, whereas
  $2.26^{+0.46}_{-0.22}$ SM background events are expected. The derived upper limit of cross section
  times branching fraction as a function of the stop mass, together with the theoretical expectation,
  is displayed in Fig.~\ref{rpvstop}. A lower limit of 155~GeV for the stop mass has been obtained,
  assuming 100\% branching fraction for the decay $\tilde{t}_1 \rightarrow b + \tau$.

  \subsection{Search for MSSM Higgs}

  In models of electroweak symmetry breaking with two Higgs doublets, such as the MSSM, there
  are five physical Higgs bosons: two neutral $CP$-even scalars, $h$ and $H$, a neutral
  $CP$-odd state, $A$, and two charged states, $H^\pm$. Neutral Higgs bosons are generically
  denoted as $\phi$. At tree level, the Higgs sector of the MSSM is fully specified by two
  parameters, generally chosen to be $M_A$, the mass of the $CP$-odd Higgs boson, and $\tan
  \beta$. At large $\tan \beta$, the coupling of the neutral Higgs bosons to down-type quarks
  and charged leptons is strongly enhanced, leading to sizeable cross sections. In this
  region, the $A$ boson is nearly degenerate in mass with the $h$ or the $H$ boson. The
  dominant decay modes are $\phi \rightarrow b \bar{b}$ ($\simeq 90$\%) and $\phi \rightarrow
  \tau^+\tau^-$ ($\simeq 8$\%), and the most promising search channels are considered to be
  either in association with $b$ quarks, $\phi\, b (\bar{b}) \rightarrow b\bar{b}\, b
  (\bar{b})$, or $\phi \rightarrow \tau^+\tau^-$.

  CDF has published results of a search for neutral supersymmetric Higgs bosons in the decay
  to $\tau^+\tau^-$ \cite{cdfhtautau}, based on 310~pb$^{-1}$ of data, and D\O\ published
  \cite{d0mssmh} results both in the $\tau^+\tau^-$ channel as well as for the decay into
  $b\bar{b}$ in association with $b$ quarks, leading to final states with three or four $b$
  jets, using integrated luminosities of 325~pb$^{-1}$ and 260~pb$^{-1}$, respectively. In
  the $\tau^+\tau^-$ channel, both experiments require one of the taus to decay
  leptonically, and the other one hadronically; in addition, D\O\ uses the $e \mu$ channel.
  The main discriminating variable in the search is the visible mass $M_{vis}$, calculated
  from the partially reconstructed tau decays. The dominant background is from $Z/\gamma^*
  \rightarrow \tau^+\tau^-$ decays.

  Searching for a MSSM Higgs in the decay into $b\bar{b}$ in association with $b$ quarks,
  the signal would show up in the dijet invariant mass distribution of the two leading
  jets. To suppress the dominant multijet background, D\O\ requires at least three jets
  with $b$ quarks identified by a secondary vertex algorithm.

\begin{figure}[bth]
    \includegraphics[height=.27\textheight]{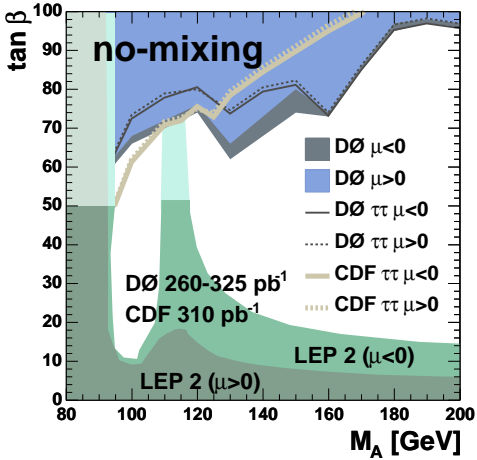}
    \caption{Excluded region in the $(M_A, \tan \beta)$ plane for the ``no-mixing'' scenario
    \cite{mssmtheo} for $\mu = - 0.2$~TeV and $\mu = + 0.2$~TeV.
    The regions labeled ``D\O\ $\mu < 0 $'' and ``D\O\ $\mu > 0$'' refer to the combination
    of the D\O\ $\tau^+\tau^-$ and multijet analyses.}
    \label{mssmhiggsfig}
  \end{figure}

  In neither of the analyses a significant access has been found, and exclusion limits
  are derived in the $(M_A, \tan \beta)$ plane. All results are summarized for one of
  the benchmark scenarios \cite{mssmtheo}, the ``no-mixing'' scenario, in
  Fig.~\ref{mssmhiggsfig}. With increasing amounts of data, these analyses impose more
  and more stringent constraints on the MSSM Higgs sector.

\section{New Gauge Bosons}

  A possible way of resolving the inherent problems of the standard model is by extending the gauge
  sector of the theory. New heavy gauge bosons are predicted in many extensions of the standard model.
  For example, in little Higgs models, the quadratically divergent radiative corrections to the Higgs
  mass are canceled individually, leading to the appearance of partners of the $W$ and $Z$ bosons at
  the TeV scale. In grand unified theories heavy partners of the electroweak bosons generally appear;
  the left-right symmetric model is a $SO(10)$ GUT extension of the SM, postulating the existence of a
  right-handed version of the weak interaction as well as an additional $Z$ boson.Finally, the sequential
  standard model, where the couplings to quarks and leptons are as in the SM, may not be gauge invariant,
  but it serves as a good benchmark for comparisons of results.

\begin{figure}[bth]
    \includegraphics[height=.25\textheight]{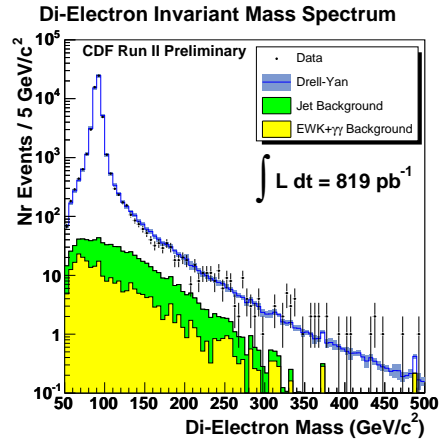}
    \caption{Dielectron invariant mass spectrum in the CDF search for electron-positron resonances.}
    \label{dielectron}
  \end{figure}

  In the search for $Z'$ bosons, the latest results are from CDF analyzing dielectron
  resonances using 819~pb$^{-1}$ of data. The invariant mass spectrum of dielectron events used in this
  search is shown in Fig.~\ref{dielectron}. No significant excess is seen at any mass value leading to a
  limit on the mass of a sequential, standard model-like $Z'$ of $m(Z')>850$~GeV at 95\% C.L. CDF also
  searches for pair production of doubly charged Higgs bosons in the lepton flavor violating modes
  $H^{++} \rightarrow e^+\tau^+$ and $H^{++} \rightarrow \mu^+\tau^+$ (and charge conjugates). At least
  three leptons are required in the final states so that the backgrounds are very small, and no events
  are observed in approximately $350$~pb$^{-1}$ of data. Interpreted in a left-right symmetric model,
  the corresponding mass limit on doubly charged Higgs bosons coupling to left-handed particles is
  $m(H^{++}) > 114 (112)$~GeV at 95\% C.L.~in the $e\tau$ ($\mu\tau$) channel.

\begin{figure}[bth]
    \includegraphics[height=.3\textheight]{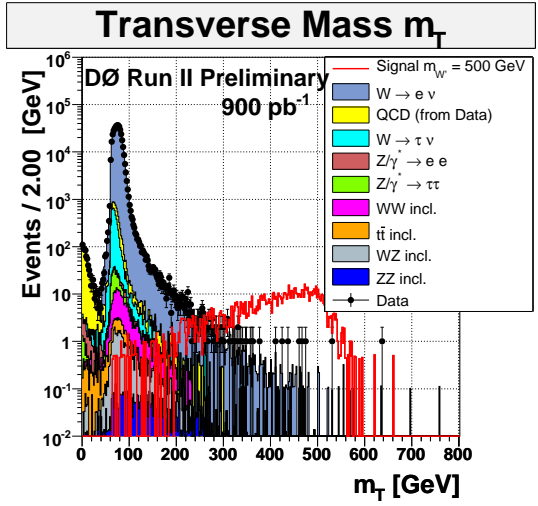}
    \caption{Transverse mass spectrum in the D\O\ search for $W'$. Also shown is the distribution for a
    hypothetical $W'$ with a mass of 500~GeV.}
    \label{wprime1}
  \end{figure}

\begin{figure}[bth]
    \includegraphics[height=.3\textheight]{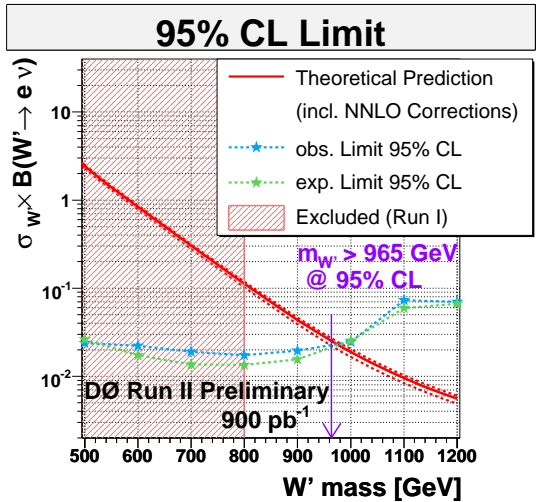}
    \caption{Cross section upper limits for the production of a SM-like $W'$
             obtained by D\O.}
    \label{wprime2}
  \end{figure}

  In the search for singly charged gauge bosons, both D\O\ and CDF looked for a SM-like $W'$ decaying
  to an electron and a neutrino. The best limit, obtained from a study of the transverse mass spectrum
  in 900~pb$^{-1}$ of data as shown in Fig.~\ref{wprime1}, is from D\O\ and requires $m(W')>965$~GeV
  at 95\% C.L. (Fig.~\ref{wprime2}).

\section{Leptoquarks}

  Leptoquarks are a natural consequence of the unification of quarks and leptons into a single
  multiplet, and as such are expected to be gauge bosons as well. While their masses may
  naturally be expected to be of the order of the unification scale, in some models they can be
  relatively light. Experimentally, it is customary to consider one leptoquark per generation.
  These are assumed to be very short-lived and decay to a quark and a lepton. The branching
  fraction to a charged lepton and a quark is then denoted as $\beta$. At hadron colliders,
  leptoquarks can be pair-produced through the strong interaction, or are singly produced. In the
  latter case the production cross section depends on the (unknown) quark-lepton coupling, which
  is generally taken to be of the same order of magnitude as the fine structure constant.

\begin{figure}[bth]
    \includegraphics[height=.2\textheight]{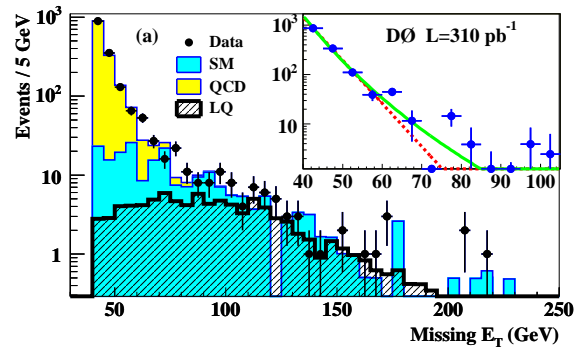}
    \caption{Missing transverse energy distribution in the D\O\ search for $\beta=0$
            leptoquarks before final cuts. The inset shows the two different fitting functions
	    used to evaluate the background from mismeasured QCD dijet events.}
    \label{lq1}
  \end{figure}

\begin{figure}[bth]
    \includegraphics[height=.2\textheight]{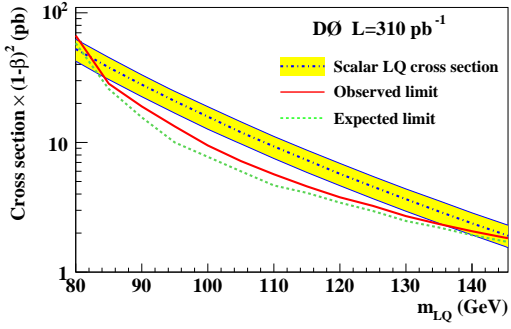}
    \caption{Cross section upper limits obtained in the search for $\beta=0$ scalar leptoquarks
             by D\O, together with the theoretical prediction.}
    \label{lq2}
  \end{figure}

  Only the three most recent results from leptoquark searches at the Tevatron are discussed
  here. D\O\ has searched for scalar leptoquarks decaying to a quark and a neutrino ($\beta=0$)
  in the jets plus missing tranverse energy topology in $310$~pb$^{-1}$ of data \cite{d0lq}.
  Experimentally this is a difficult analysis which suffers from substantial QCD dijet
  background due to mismeasured jets. To mitigate this, D\O\ requires exactly two acoplanar
  jets. The ensuing missing transverse energy distribution, before final analysis cuts, is shown
  in Fig.~\ref{lq1}. The background from QCD dijet events, dominant at low missing transverse
  energy, is extrapolated to higher values using two different fitting functions as shown in the
  inset. The dominant non-QCD standard model background is $Z$ boson plus jets production with
  the $Z$ decaying to a pair of neutrinos. No excess is observed, so D\O\ sets a limit  on the
  leptoquark mass of $M_{LQ} > 136$~GeV at 95\% C.L. (Fig.~\ref{lq2}).

\begin{figure}[bth]
    \includegraphics[height=.25\textheight]{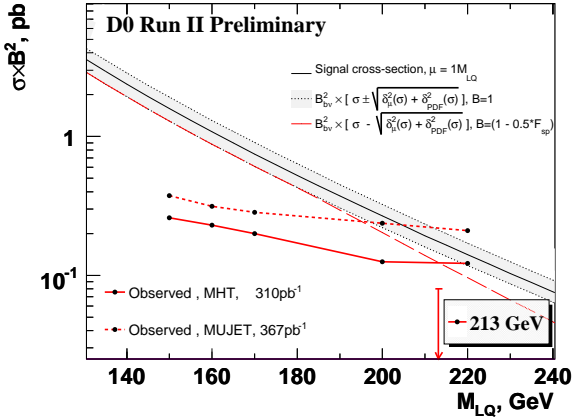}
    \caption{Cross section upper limits by D\O\ obtained in the search for $\beta=0$ scalar leptoquarks
             of the third generation, together with the theoretical prediction.}
    \label{lq3}
  \end{figure}

  D\O\ searched for the pair production of scalar third generation leptoquarks decaying into a
  tau neutrino and a $b$ quark, leading to a final state of two $b$ jets and \MET, identical to
  the sbottom search described earlier. Both jets are required to be identified as $b$ jets
  using a lifetime based algorithm. Further cuts on \MET\ and $H_T$ are optimized depending on
  the leptoquark mass $M_{LQ}$. For $M_{LQ} = 220$~GeV, $\MET > 90$~GeV and $H_T >190$~GeV, one
  event is selected in the data corresponding to an integrated luminosity of 310~pb$^{-1}$,
  while $2.6 \pm 0.6$ events are expected from SM backgrounds. In the absence of an excess in
  the data, a limit of $M_{LQ} > 219$~GeV at 95\% C.L. is set, assuming a branching fraction $B
  ( LQ \rightarrow b \nu ) = 1$; allowing also decays to $\tau t$, the limit is reduced to
  $M_{LQ} > 213$~GeV (Fig.~\ref{lq3}).

  CDF has released results on a search for vector leptoquarks of the third generation in
  the $LQ_3 \rightarrow b \tau$ decay channel. The signature consists of a dijet plus
  ditau final state, in which one tau is required to decay leptonically and the other
  hadronically. The main discriminating variables are the number of jets and an $H_T$-like
  variable, the scalar sum of the transverse energies of all jets, leptons and \MET. This
  allows CDF to set a limit at $M_{LQ} > 294$~GeV assuming $\beta=1$ and using
  322~pb$^{-1}$ of data. Note that this limit is higher than the typical limits on
  leptoquark masses at the Tevatron due to the model choice of vector leptoquarks, which
  have a much larger production cross section than the scalar leptoquarks which are
  usually chosen.

\section{Large Extra Dimensions}

\begin{figure}[bth]
    \includegraphics[height=.2\textheight]{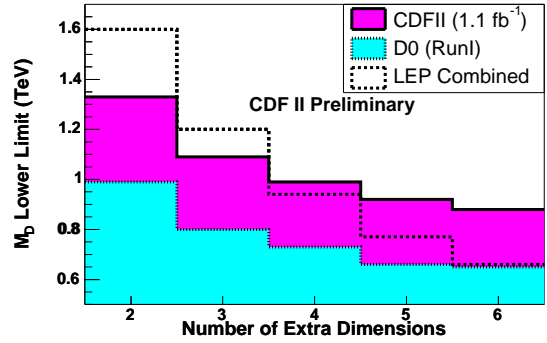}
    \caption{CDF limits on the effective Planck scale $M_D$ as a function of the number
    of extra dimensions, compared to previous results.}
    \label{ledfig}
  \end{figure}

  Models postulating the existence of extra spatial dimensions have been proposed to solve the
  hierarchy problem posed by the large difference between the Planck scale $M_{pl} \simeq
  10^{16}$~TeV, at which gravity is expected to become strong, and the scale of electroweak
  symmetry breaking, $\simeq 1$~TeV. In the original large extra dimensions model of
  Arkani-Hamed, Dimopoulos and Dvali \cite{add}, in which only gravitons propagate in the bulk
  but all standard model fields are confined to a 3-brane, a tower of Kaluza-Klein excitations
  of the graviton emerges. The graviton states are too close in mass to be distinguished
  individually, and the coupling remains small, but the number of accessible states is very
  large. It is therefore possible to produce gravitons which immediately disappear into bulk
  space, leading to an excess of events with a high transverse energy jet and large missing
  transverse energy, the monojet signature: $q\bar{q} \rightarrow g G$, $qg \rightarrow qG$
  and $gg \rightarrow gG$, where $G$ is the emitted graviton. The dominant standard model
  backgrounds are the production of $Z$ or $W$ bosons plus jets, with the $Z$ decaying to a
  pair of neutrinos or the lepton from the $W$ decay escaping detection. Using 1.1~fb$^{-1}$
  of data, CDF has recently updated their published \cite{cdfled} analysis, and set limits on
  the effective Planck scale between $M_D > 1.33$~TeV and $M_D > 0.88$~TeV for a number of
  extra dimensions ranging from 2 to 6 (Fig.~\ref{ledfig}).

  In the model by Randall and Sundrum \cite{rs} gravity is located on a $(3+1)$-dimensional
  brane, the Planck brane, that is separated from the standard model brane in a fifth
  dimension with warped metric. In the simplest version of this model gravitons are the only
  particles that can propagate in the extra dimension. The gravitons appear as towers of
  Kaluza-Klein excitations with masses and widths determined by the parameters of the model.
  These parameters can be expressed in terms of the mass of the first excited mode of the
  graviton, $M_1$, and the dimensionless coupling to the standard model fields,
  $k\sqrt{8\pi}/M_{pl}$. If it is light enough, the first excited graviton mode could be
  resonantly produced at the Tevatron. It is expected to decay to fermion-antifermion and to
  diboson pairs.

\begin{figure}[bth]
    \includegraphics[height=.3\textheight,angle=90]{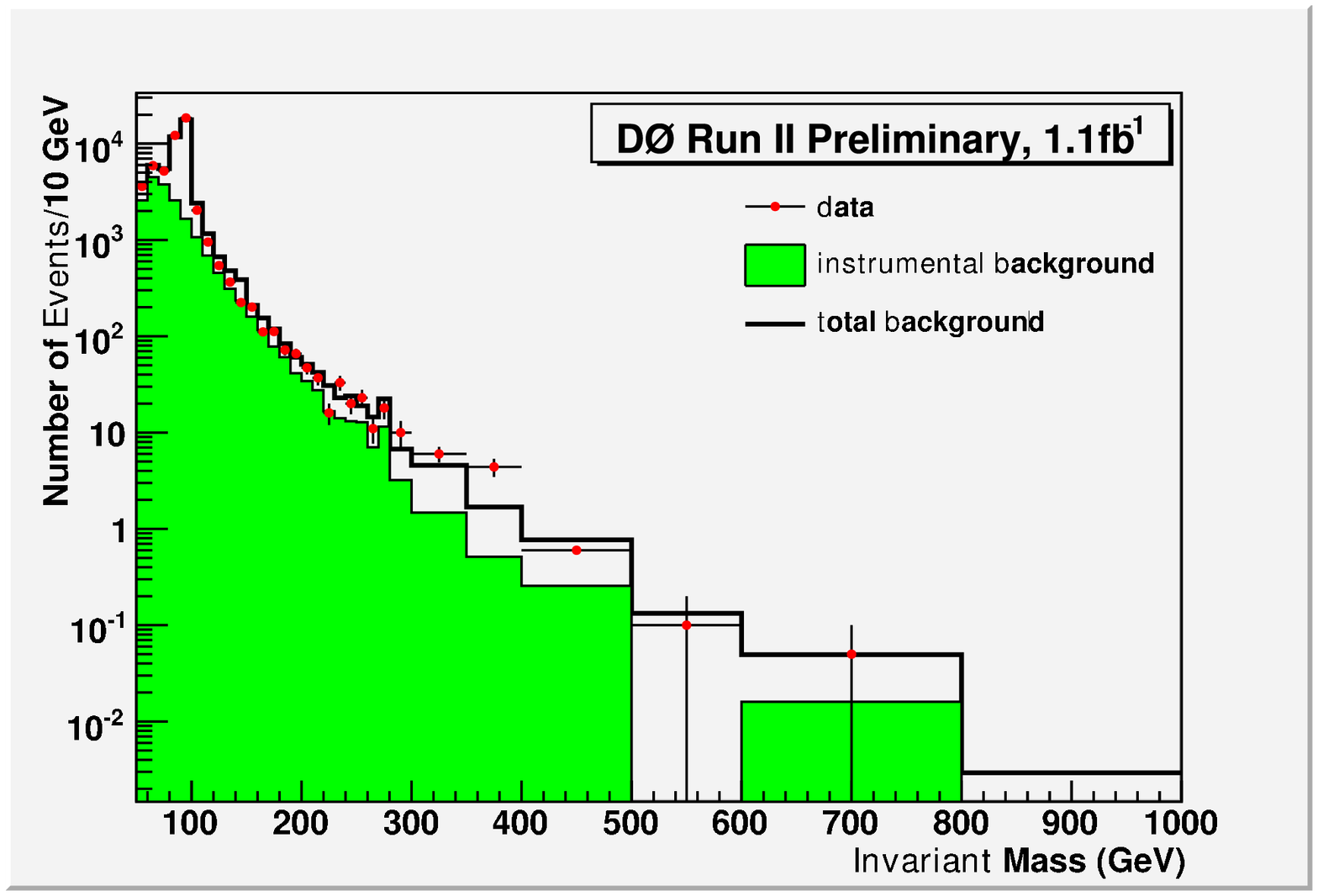}
    \caption{Invariant mass spectrum of diphotons and dielectrons observed by D\O\ in the search for
      Randall-Sundrum gravitons.}
    \label{rs1}
  \end{figure}

\begin{figure}[bth]
    \includegraphics[height=.26\textheight]{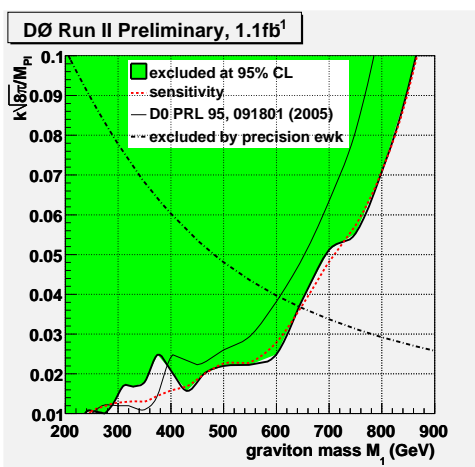}
    \caption{Excluded region in the plane of $k\sqrt{8\pi}/M_{pl}$ and graviton mass by D\O.}
    \label{rs2}
  \end{figure}

  Both CDF and D\O\ have recently presented new results in the search for Randall-Sundrum
  gravitons based on $0.8-1.2$~fb$^{-1}$ (CDF) and 1.1~fb$^{-1}$ (D\O) of data,
  respectively. Both experiments analyze the $e^+e^-$ and $\gamma\gamma$ final states. The
  invariant mass spectrum measured by D\O\ is shown in Fig.~\ref{rs1}, and general agreement
  between data and the background expectation is observed. Using a sliding mass window,
  upper cross section limits are derived, which are then translated into lower mass limits
  for the lowest excited mode of Randall-Sundrum gravitons (Fig.~\ref{rs2}). For a coupling
  parameter $k\sqrt{8\pi}/M_{pl} = 0.1$ ($0.01$), masses $M_1 < 865$ ($240$)~GeV are excluded at
  the 95\% C.L. by the D\O\ analysis. The corresponding CDF exclusion limits are
  $M_1 < 875$ ($242$)~GeV.

\section{Excited Fermions and Other Resonances}

\begin{figure}[bth]
    \includegraphics[height=.25\textheight]{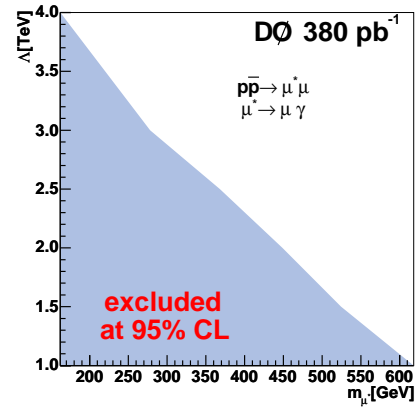}
    \caption{Exclusion region from the D\O\ search for excited muons produced in contact interactions
    in the decay mode $\mu^* \rightarrow \mu \gamma$.}
    \label{mustar}
  \end{figure}

  In the search for excited leptons, CDF has published previously \cite{cdfestar} results of a
  search for excited electrons. Recently, both experiments have presented their analyses of
  excited muon production \cite{d0mustar,cdfmustar}. They searched for associated
  production of a muon and an excited muon, with the latter decaying to a muon and a photon.
  The production is approximated as a contact interaction, while the decay is assumed to
  proceed either exclusively through a gauge interaction (CDF) or a combination of gauge and
  contact interactions, with the relative fraction of the two depending on the mass of the
  excited muon and the compositeness scale $\Lambda$ (D\O). Both experiments obtain very
  similar results using 371 (CDF) and 380 (D\O) pb$^{-1}$ of data. The D\O\ result is shown in
  Fig.~\ref{mustar}, excluding for example $m_{\mu^*} < 618$~GeV for $\Lambda = 1$~TeV at 95\%
  C.L. To make a comparison with LEP and HERA results easier, CDF also reinterprets the result
  in a gauge mediated model with Drell-Yan-like production of the $\mu\mu^*$ pair with coupling
  $f/\Lambda$.

  Both CDF and D\O\ have presented recent results on the search for resonances decaying to
  $\mbox{dileptons}+X$. CDF has performed two analyses, with the first using $H_T$ as its main
  discriminating variable. After selecting events with two high $p_T$ leptons ($e$, $\mu$), a
  control region with $H_T < 200$~GeV is used to establish a good understanding of the data,
  and the signal region is chosen to have $H_T > 400$~GeV, including at least two jets each
  with transverse energy $E_T>50$~GeV. Using 929~pb$^{-1}$ of data, two events are observed in
  the signal region, with an expected background of $2.9 \pm 1.5$ events. A limit of
  $\sigma<0.4$~pb at 90\% C.L. is set on the production cross section of right-handed down-type
  quarks with a mass of 300~GeV as proposed in \cite{bpt}.

\begin{figure}[bth] \setlength{\unitlength}{1cm}
    \begin{picture}(8.6,8.0)(0.0,0.0)
      \put(-2.0,8.5) {\includegraphics[height=.32\textheight,angle=270]{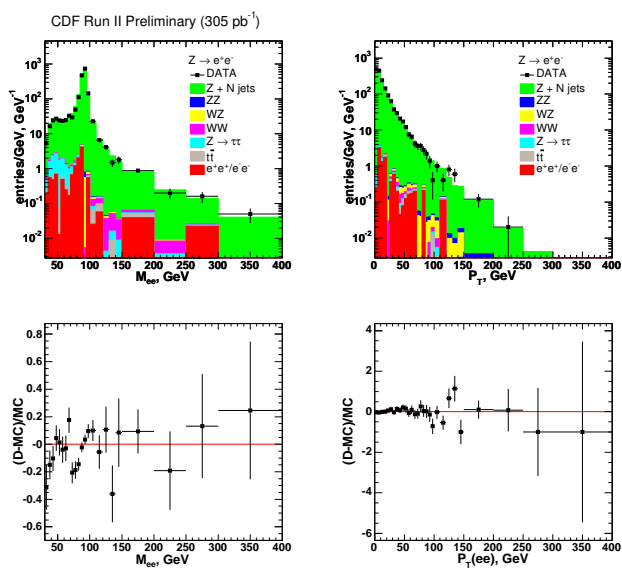}}
    \end{picture}
    \caption{Invariant mass and transverse momentum distribution of electron-positron pairs measured by CDF in
            305~pb$^{-1}$ of data; a mass cut $66 < M_{ee} < 116$~GeV has been applied for the right hand figures.}
    \label{zpt}
  \end{figure}

  In a second analysis, CDF studied the transverse momentum distribution of $Z$ bosons. This has
  the advantage of being insensitive to the nature of the other decay products, but it is of
  course more difficult and potentially less sensitive than a direct resonance search. Selecting
  events with a dielectron invariant mass compatible with the mass of the $Z$ boson,
  $66 < M_{ee} < 116$~GeV, CDF measured the $Z$ transverse momentum distribution shown in
  Fig.~\ref{zpt}. From this they determined an upper limit on the anomalous production of $Z$ bosons
  as a function of transverse momentum using 305~pb$^{-1}$ of data. The 95\% C.L. limit ranges
  from about 1~pb for $p_T(Z)=20$~GeV to approximately 2~fb for $p_T(Z)=200$~GeV. A similar search
  was carried out in the $Z\rightarrow \mu^+\mu^-$ channel.

\begin{figure}[bth]
    \includegraphics[height=.25\textheight]{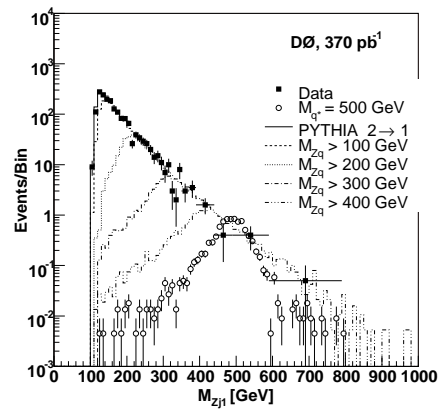}
    \caption{Invariant mass of the Z boson and leading jet as measured by D\O\ in 370~pb$^{-1}$
            of data. The curves with different $M_{Zq}$ thresholds correspond to different generation
	    thresholds for $2\rightarrow 2$ processes in Pythia, and the open circles represent the signal due
            to an excited quark of mass $m_{q^*} = 500$~GeV and narrow width.}
    \label{qstar}
  \end{figure}

  D\O\ explicitly searched for a $Z$ boson plus jet resonance by combining the $Z$ boson plus jet mass
  spectrum and the $Z$ boson transverse momentum distribution as discriminating variables \cite{qstarpub}.
  The invariant mass distribution, using the decay $Z \rightarrow e^+e^-$ and events with $80 < M_{ee} <
  102$~GeV, measured in 370~pb$^{-1}$ of data is shown in Fig.~\ref{qstar}. No excess is observed and a
  limit is set on the mass of an excited quark as proposed in \cite{excited} at $M_{q^*} > 520$~GeV at
  95\% C.L.

\section{Events with Leptons and Photons}

  In Run I, CDF reported an excess of events with a photon, a lepton and large missing
  transverse energy compared to standard model expectations \cite{cdf2002}. This excess,
  corresponding to a $2.7$ sigma effect, was deemed ``an interesting result'', but not ``a
  compelling observation of new physics''. CDF has repeated this analysis in Run II using
  305~pb$^{-1}$ of data \cite{cdf2006}. The data is now compatible with standard model
  expectations in all channels, suggesting that the excess observed in Run I was due to a
  statistical fluctuation.

  In a similar spirit, CDF searches for events with two photons and an additional object,
  which can be an electron, a muon, a third photon, or \MET. The analysis is based on
  $1.0 - 1.2$~fb$^{-1}$ of data. In the $\gamma\gamma +$ \MET\ search, a sliding cut on
  \MET\ from 20 to $> 150$~GeV is used, and for all cut values agreement between the data
  and the background expectation is found. In the $e\gamma\gamma$ and $\mu\gamma\gamma$
  final states, 1 and 0 events are selected in the data, while $3.79 \pm 0.54$ and $0.71
  \pm 0.10$ events are expected from SM backgrounds. Finally, 4 events are found with
  three photons, with a background expectation of $2.2 \pm 0.6$. Unfortunately, no excess
  is observed over expectations.

\section{Standard Model Higgs}

  The search for the SM Higgs is the focus of much attention both inside and outside the
  Tevatron community. Therefore its current state is briefly mentioned here.

\begin{figure}[bth]
    \includegraphics[height=.25\textheight]{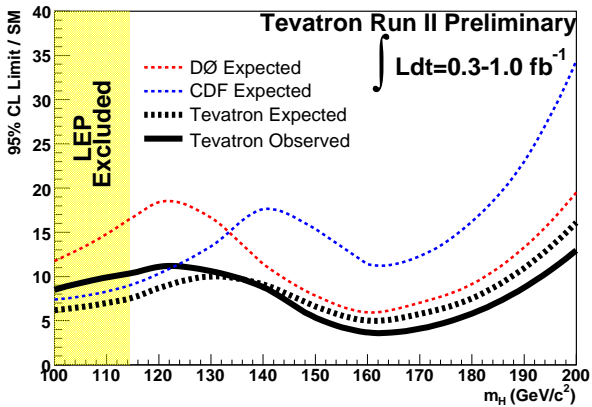}
    \caption{Expected and observed lower limits on the Higgs boson production cross section relative to
       the SM expectation, as a function of the Higgs mass. In addition to the combined result, the
       expected 95\% C.L.~ratios for the CDF and D\O\ experiments alone are shown.}
    \label{higgs1}
  \end{figure}

\begin{figure}[bth]
    \includegraphics[height=.22\textheight]{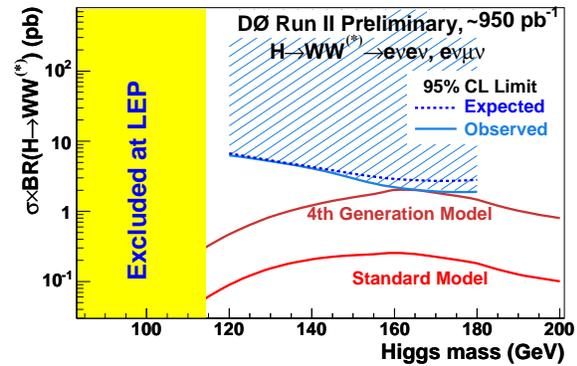}
    \caption{Excluded cross section times branching ratio at 95\% C.L.~in the D\O\ $H \rightarrow W^+ W^-$
     analysis together with expectations from standard model Higgs boson production and an alternative model
     with four fermion generations.}
    \label{higgs2}
  \end{figure}

  The available results of SM Higgs searches at the Tevatron have recently been combined for
  the first time \cite{higgscomb}. The results are for data corresponding to integrated
  luminosities ranging from 360 -- 1000~pb$^{-1}$ at CDF and 260 -- 950~pb$^{-1}$ at D\O. These
  searches are for SM Higgs bosons produced in association with vector bosons ($p\bar{p}
  \rightarrow W/Z\, H \rightarrow l \nu b \bar{b} / \nu \bar{\nu} b \bar{b} / l^+l^- b \bar{b}$
  or $p\bar{p} \rightarrow W H \rightarrow W W^+ W^-$) or singly through gluon-gluon fusion
  ($p\bar{p} \rightarrow H \rightarrow W^+ W^-$), separated into 16 mutually exclusive final
  states. Special care has been taken to account for systematic uncertainties and their
  correlations, and to ensure a statistically robust result. The combined result is shown in
  Fig.~\ref{higgs1} in terms of the ratio of limits set to the SM cross sections as a
  function of Higgs mass. A value of $< 1$ would indicate a Higgs mass excluded at 95\% C.L.
  The observed limits are factors of 10.4 and 3.8 of the SM expectation at Higgs masses of
  $m_H = 115$~GeV and $m_H = 160$~GeV, respectively. These results represent an improvement
  in search sensitivity over those obtained for individual experiments, as can also be seen
  from Fig.~\ref{higgs1}. With the expected increase in integrated luminosity as well as
  improvements in analysis techniques, the Tevatron experiments have a realistic chance to
  reach sensitivity for a SM Higgs before the LHC delivers first results.
  Already now, the results are placing constraints on certain scenarios beyond the standard
  model, for example in models with four or more fermion generations \cite{fourgen} (Fig.~\ref{higgs2}).

\section{Conclusions}

  Both the Tevatron and HERA continue their good performance, and the D\O, CDF, H1 and ZEUS experiments
  collect large amounts of data and search for new physics in many channels. Only recent results have
  been reported here, and many other analyses are in progress. From HERA, the most interesting results
  continue to be in the area of isolated leptons. At the Tevatron, many different models are tested
  with ever increasing sensitivity, but no discoveries are to be reported yet. So far, new physics
  has remained hidden. The search for the SM Higgs continues with high priority, and a substantial amount
  of data is yet to be recorded and analyzed before the LHC can be expected to start delivering results.

\bigskip

\begin{acknowledgments}

  I would like to thank the organizers for a well organized conference with an exciting
  programme in a beautiful location. Many thanks to my colleagues from the D\O, CDF,
  H1 and ZEUS collaborations for their help in preparing this talk.

\end{acknowledgments}

\end{document}